\documentclass[aps,pra,reprint,superscriptaddress,amsmath,amssymb,floatfix,showkeys,longbibliography]{revtex4-2}
\usepackage{graphicx}
\usepackage{mathtools}
\usepackage{booktabs}
\usepackage{microtype}
\usepackage{braket}
\usepackage{dsfont}
\usepackage[colorlinks=true,allcolors=blue]{hyperref}
\DeclareMathOperator{\Tr}{Tr}
\newcommand{\1}{\mathds{1}}
\newcommand{\ee}{\mathcal{E}}
\newcommand{\capc}{\mathcal{C}}
\newtheorem{proposition}{Proposition}

\begin{document}
\title{Entangled Measurement Outcomes for Optimal Postselection in Quantum Batteries}
\author{Yali Du}
\email{Duyali_123@126.com}
\affiliation{Faculty of Mathematics and Statistics, Hubei University, Wuhan, Hubei 430062, China}
\author{Zhipei Zhou}
\email{zppei2022@163.com}
\affiliation{School of Mathematics and Statistics, Central China Normal University, Wuhan, Hubei 430079, China}
\author{Xiaoli Hu}
\email[Corresponding author: ]{xiaolihumath@163.com}
\thanks{ORCID: \href{https://orcid.org/0000-0001-5462-9865}{0000-0001-5462-9865}.}
\affiliation{School of Artificial Intelligence, Jianghan University, Wuhan, Hubei 430056, China}

\begin{abstract}
Postselection can enhance the performance of a selected quantum battery branch, raising the question of whether the optimum can always be achieved by product measurements on the auxiliary subsystems. 
For battery auxiliary states obtained by mixing a pure state with global white noise of strength $p$, the conditional ergotropy reduces to a generalized Rayleigh quotient. 
For $0<p<1$, product postselection attains the joint optimum if and only if the eigenspace associated with the largest generalized eigenvalue contains a product vector.
For two auxiliary qubits, a strict product–joint gap therefore occurs precisely when this eigenspace is one dimensional and spanned by an entangled vector. 
Noisy W and GHZ states display qualitatively different behavior: the three qubit W state exhibits a strict gap throughout $0<p<1$, whereas the three qubit GHZ state admits a product optimum. 
For interacting two qubit batteries, the optimal postselection can switch from a Bell outcome to a product outcome as the coupling strength is varied. 
By contrast, the outcome averaged capacity and ergotropy are independent of the measurement basis, showing that the joint measurement advantage is genuinely postselective.
\end{abstract}

\maketitle

\section{Introduction}\label{sec:introduction}

Measurements on auxiliary systems correlated with a quantum battery can conditionally prepare battery states with outcome dependent energetic properties. 
We ask whether the optimal branch can be reached by product postselection on the auxiliaries or instead requires an entangled joint measurement outcome.
Ergotropy quantifies the maximum work extractable from a quantum state by cyclic unitary operations \cite{Allahverdyan2004}, whereas battery capacity is the energy range accessible along its unitary orbit \cite{Yang2023}. Capacity is determined solely by the spectra of the state and Hamiltonian, while ergotropy also depends on the state's mean energy and therefore on its orientation relative to the energy eigenbasis \cite{Francica2020,Shi2022}. Consequently, a measurement that is optimal for one quantity need not be optimal for the other.

Correlations and measurement based control can substantially modify the energy stored in and extractable from quantum batteries \cite{Alicki2013,Campaioli2024}.
Measurement assisted work extraction is described by daemonic ergotropy, which has been extended to generalized measurements and multipartite settings \cite{Francica2017,Bernards2019}. 
Related protocols use auxiliary measurements to distill energy \cite{Chaki2025} or enhance battery capacity through projective, generalized, and weak measurements \cite{Zhang2024,Wang2025,Zheng2026}. 
More recently, energy distillation protocols have been studied under both unrestricted joint projective measurements and experimentally limited measurement settings \cite{Chaki2026}. 
In particular, Ref.~\cite{Wang2025} investigated noisy three qubit W and GHZ batteries under local projective measurements.
We address a different question: whether the optimum over arbitrary joint projections for a selected battery branch can already be reached by a product measurement outcome. 
We therefore optimize over all product vectors and compare the result with the unrestricted joint optimum. 
This setting is distinct from measurement based charging, stabilization, and feedback protocols \cite{Yan2023,Gherardini2020,Mitchison2021,He2026QIP}, as well as from nonselective measurements performed directly on the battery \cite{Du2025}. 
Here, the measurement acts only on the auxiliary subsystem and the battery state is conditioned on a selected outcome. 
We consider both the resulting conditional energetic quantities and their Born probability weighted contributions. 
These quantities do not include the energetic costs of state preparation, measurement, or reset, and should therefore not be interpreted as the net work of a full measurement and feedback cycle \cite{Rogers2022}.

For a pure battery auxiliary state admixed with global white noise, every conditional battery state consists of an isotropic background and a rank-one contribution. 
For a fixed battery Hamiltonian, the conditional capacity depends only on the weight of this rank-one term, whereas the ergotropy also depends on the energy expectation of its pure state component.
For $0<p<1$, optimizing the conditional ergotropy reduces to a generalized eigenvalue problem.
A product outcome attains the unrestricted joint optimum if and only if the eigenspace associated with the largest generalized eigenvalue contains a product vector. 
For two auxiliary qubits, the product and joint optima differ precisely when this eigenspace is one-dimensional and spanned by an entangled vector.

The four examples realize distinct regimes of conditional ergotropy for $0<p<1$. 
An entangled outcome is necessary to attain the joint optimum for W$_3$, whereas a product outcome suffices for GHZ$_3$. 
For GHZ$_4$, any nonzero coupling $J$ makes an entangled outcome necessary to attain the joint optimum; for W$_4$, the optimizer changes from a Bell vector to $|00\rangle$ as $J$ crosses $\epsilon_A+\epsilon_B$. 
In the four qubit GHZ case, the exact optimization over product vectors also shows that the conditional ergotropy and its success probability weighted contribution can favor different product measurement outcomes.
Despite these differences at the level of individual outcomes, the capacity and ergotropy averaged over all outcomes are independent of the measurement basis for the global white noise family considered here. 
The advantage of a joint measurement therefore lies in how these fixed averages are redistributed among branches, and can enhance the contribution carried by a selected outcome after weighting by its success probability. 
Section~\ref{sec:protocol} develops the general optimization framework, while Secs.~\ref{sec:three} and \ref{sec:four} apply it to noisy W and GHZ batteries. The derivations, constrained optimizations, and boundary cases are collected in Appendix~\ref{app:optimization}.


\section{Postselected projective measurements}\label{sec:protocol}


\subsection{Measurement classes and conditional states}

Let $\rho_{TM}$ denote the joint state of a target battery $T$ of dimension $d_T$, with Hamiltonian $H_T$, and an auxiliary subsystem $M$ of dimension $d_M$.
The symbol $\1$ denotes the identity operator, with its subscript specifying the subsystem or the dimension of the Hilbert space.
A complete rank one projective measurement on $M$ is specified by an orthonormal basis $\{|v_i\rangle\}_{i=1}^{d_M}$ and the projectors
\begin{align}
K_i\coloneqq|v_i\rangle\langle v_i|,\quad
 \langle v_i|v_j\rangle=\delta_{ij},\quad
 \sum_{i=1}^{d_M}K_i=\1_M.
\label{eq:projectors}
\end{align}
The Born probability and normalized conditional state for outcome $i$ are
\begin{align}
 P_i&\coloneqq\Tr \left[(\1_T\otimes K_i)\rho_{TM}\right],\label{eq:probability}\\
 \rho_T^i&\coloneqq\frac{1}{P_i}\Tr_M \bigl[(\1_T\otimes K_i)\rho_{TM} (\1_T\otimes K_i)\bigr].
\label{eq:conditional}
\end{align}
Here, $P_i$ is the Born probability of outcome $i$, and the normalized conditional state $\rho_T^i$ is defined whenever $P_i>0$.

In the examples below, the auxiliary subsystem $M=M_1M_2$ consists of two qubits.
Using the ordered basis $(|e_1\rangle,|e_2\rangle,|e_3\rangle,|e_4\rangle)\coloneqq(|00\rangle,|01\rangle,|10\rangle,|11\rangle)$, we compare three classes of complete projective measurements:

\begin{list}{}{\setlength{\leftmargin}{0pt}\setlength{\itemindent}{0pt}\setlength{\itemsep}{0.5em}\setlength{\parsep}{0pt}\setlength{\topsep}{0.4em}}
\item \textbf{Protocol I (Common Local Basis).}
The same local unitary $U\in\mathrm{U}(2)$ defines the measurement basis on both auxiliary qubits:
\begin{align}
 |v_{rs}^{\mathrm I}\rangle
 \coloneqq U|r\rangle\otimes U|s\rangle, \quad r,s\in\{0,1\}.
\end{align}

\item \textbf{Protocol II (Independent Local Bases).}
The two local bases are defined by independently chosen unitaries $U_1,U_2\in\mathrm{U}(2)$:
\begin{align}
 |v_{rs}^{\mathrm{II}}\rangle\coloneqq U_1|r\rangle\otimes U_2|s\rangle, \quad r,s\in\{0,1\}.
\end{align}

\item \textbf{Protocol III (Joint Basis).}
An arbitrary unitary $V\in\mathrm{U}(4)$ defines a joint basis of the auxiliary pair:
\begin{align}
 |v_i^{\mathrm{III}}\rangle\coloneqq V|e_i\rangle,\quad i=1,\ldots,4.
\end{align}
\end{list}
These classes satisfy
$\mathrm{I}\subset\mathrm{II}\subset\mathrm{III}$,
and a basis in Protocol III may contain both product and entangled vectors.
Throughout, ``joint'' refers only to the auxiliary bipartition $M_1|M_2$; no measurement is performed on the target.

For a selected outcome, Protocol I restricts the two local factors to be identical or orthogonal up to phase, whereas Protocol II allows an arbitrary product vector.
We therefore use Protocol II as the benchmark for product measurements and retain Protocol I only as a more restricted reference.
At the level of a selected branch, Protocol II exhausts all product vectors, although it does not include every complete product basis or more general adaptive measurements using local operations and classical communication (LOCC).

Write the joint basis as
\begin{align}
 |v_i\rangle=\sum_{k=1}^{4}c_{ki}|e_k\rangle, \quad
 \sum_{k=1}^{4}\overline{c_{ki}}c_{kj}=\delta_{ij}.
 \label{eq:measurementmatrix}
\end{align}
Here $c_{ki}$ are the complex amplitudes of $|v_i\rangle$ in the ordered auxiliary basis, and the overbar denotes complex conjugation.
For the selected outcome, we relabel the corresponding vector as
\begin{align}
 |v\rangle=c_{11}|00\rangle+c_{21}|01\rangle+c_{31}|10\rangle+c_{41}|11\rangle,
 \label{eq:v}
\end{align}
with $\sum_{k=1}^{4}|c_{k1}|^2=1$.
The vector $|v\rangle$ is a product vector if and only if
$c_{11}c_{41}=c_{21}c_{31}$.
Any normalized vector can be completed to an orthonormal basis, so Eq.~\eqref{eq:v} covers every possible selected outcome in Protocol III.


\subsection{Capacity and ergotropy}

Let $\lambda_0\leqslant\cdots\leqslant\lambda_{d-1}$ be the eigenvalues of a $d$-dimensional state $\rho$, and let $E_0\leqslant\cdots\leqslant E_{d-1}$ be the eigenvalues of its Hamiltonian $H$. 
The battery capacity is \cite{Yang2023}
\begin{align}
	\mathcal C(\rho;H)
	\coloneqq
	\sum_{k=0}^{d-1}\lambda_k \left(E_k-E_{d-1-k}\right),
	\label{eq:capacity}
\end{align}
whereas the ergotropy is \cite{Allahverdyan2004}
\begin{align}
	\mathcal E(\rho;H)
	\coloneqq
	\Tr(\rho H)-\sum_{k=0}^{d-1}\lambda_k E_{d-1-k}.
	\label{eq:ergotropy}
\end{align}
The passive rearrangement assigns the largest populations to the lowest energy levels, while the active rearrangement reverses this ordering.
Accordingly, $\mathcal C$ is the energy difference between the active and passive rearrangements, whereas $\mathcal E$ is the difference between the actual mean energy and the passive energy.

For outcome $i$, we also introduce the branch contributions weighted by the outcome probability $P_i$
\begin{align}
Y_{\mathcal C,i}&\coloneqq P_i\,\mathcal C(\rho_T^i;H_T),\\
 Y_{\mathcal E,i}&\coloneqq P_i\,\mathcal E(\rho_T^i;H_T).
\label{eq:weighted_quantities}
\end{align}
Summed over a complete measurement, these quantities define the capacity and ergotropy averaged over outcomes.
The ergotropy average is the daemonic ergotropy associated with the chosen measurement~\cite{Francica2017,Bernards2019}.
For a selected measurement vector $|v\rangle$, let $P_v$ denote its Born probability and $\rho_T^v$ the corresponding conditional state.
Within a given measurement class, we call $|v\rangle$ a simultaneous optimizer if it attains the separate maxima of $P_v,\, \mathcal C(\rho_T^v;H_T)$ and
$\mathcal E(\rho_T^v;H_T)$.
No optimality condition is imposed on the remaining outcomes of the measurement basis.

Throughout, the resource state is treated as supplied, and work extraction is allowed to use arbitrary cyclic unitary control of the entire target, including both battery qubits in Sec.~\ref{sec:four}.
Preparation, measurement, and reset costs are not included.
Accordingly, $Y_{\mathcal E,v}$ is a contribution of a selected branch weighted by its probability, not the net work of a complete measurement and feedback cycle \cite{Rogers2022,Manzano2018}.


\subsection{Conditional spectra and product attainability}\label{subsec:criterion}

We consider the global white noise family
\begin{align}
	\rho_{TM}
	\coloneqq
	(1-p)|\Psi\rangle\langle\Psi|
	+\frac{p}{d_Td_M}\1_{TM},
	\quad
	0\leqslant p\leqslant1,
	\label{eq:generalNoise}
\end{align}
where $|\Psi\rangle$ is a pure state on $\mathcal H_T\otimes\mathcal H_M$, and $p$ is the white noise strength.
For this family, every conditional target state consists of an isotropic background plus a rank-one contribution. 
This simple spectral structure allows the product attainability problem to be treated exactly and also leads to closed expressions for the complete outcome averages derived below.

For a measurement outcome $i$, define
\begin{align}
	\ket{\chi_i}
	\coloneqq
	(\1_T\otimes\bra{v_i})\ket{\Psi}.
\end{align}
We also denote by
\begin{align}
	\rho_T^\Psi
	\coloneqq
	\Tr_M \left[\ket{\Psi}\bra{\Psi}\right]
\end{align}
the reduced target state of the pure resource. 
The unnormalized conditional state and the corresponding outcome probability are then
\begin{align}
	P_i\rho_T^i
	&=
	\frac{p}{d_Td_M}\1_T
	+(1-p)|\chi_i\rangle\langle\chi_i|,
	\label{eq:generalBranch}\\
	P_i
	&=
	\frac{p}{d_M}
	+(1-p)n_i.
\end{align}
Here, $n_i\coloneqq
\braket{\chi_i|\chi_i}$ is the probability of outcome $i$ for the pure state $|\Psi\rangle$. 
Hence the conditional spectrum contains a uniform background, with only one eigenvalue shifted by the rank-one term.

Let $E_{\min}$ and $E_{\max}$ denote the minimum and maximum eigenvalues of $H_T$, respectively. 
The probability weighted capacity and ergotropy of the conditional state $\rho_T^i$ are then
\begin{align}
	P_i\,\capc(\rho_T^i;H_T)
	&=(1-p)n_i\bigl(E_{\max}-E_{\min}\bigr),
	\label{eq:generalWeightedC}\\
	P_i\,\ee(\rho_T^i;H_T)
	&=(1-p)
	\langle\chi_i|
	\bigl(H_T-E_{\min}\1_T\bigr)
	|\chi_i\rangle.
	\label{eq:generalWeightedE}
\end{align}
For $0<p<1$ and $E_{\max}>E_{\min}$, both the success probability $P_i$ and the conditional capacity increase monotonically with $n_i$. 
The ergotropy also depends on the energy expectation of the normalized pure component $|\chi_i\rangle/\sqrt{n_i}$ whenever $n_i>0$. 
At $p=0$, every conditional state with nonzero probability is pure and therefore has the same capacity, while its ergotropy can still vary with the measurement outcome.

Let the auxiliary system decompose as $M=M_1M_2$, with both local Hilbert spaces finite dimensional.
Choose fixed orthonormal bases $\{\ket{\alpha}_T\}$ and $\{\ket{\mu}_M\}$, where $\{\ket{\mu}_M\}$ is taken to be a product basis with respect to the bipartition $M_1|M_2$.
We write
\begin{align}
	\ket{\Psi}
	=
	\sum_{\alpha,\mu}
	F_{\alpha\mu}\,
	\ket{\alpha}_T\ket{\mu}_M,
\end{align}
where
$F\coloneqq(F_{\alpha\mu})\in\mathbb C^{d_T\times d_M}$
is the coefficient matrix of $\ket{\Psi}$ with respect to the bipartition $T|M$.

For a measurement vector $\ket v=\sum_\mu c_\mu\ket{\mu}_M$, let $c\coloneqq(c_\mu)_\mu$ denote its coefficient column. 
The corresponding projected pure component on the battery is then
\begin{align}
	\ket{\chi_v}
	=
	(\1_T\otimes\bra v)\ket{\Psi}
	=
	F\overline c.
\end{align}
Setting $z\coloneqq\overline c$, define
\begin{align}
	R&\coloneqq F^\dagger F,\\
	Q&\coloneqq F^\dagger
	\bigl(H_T-E_{\min}\1_T\bigr)F,\\
	G_p&\coloneqq \frac{p}{d_M}\1_M+(1-p)R.
	\label{eq:criterionOperators}
\end{align}
For $0<p<1$, $G_p$ is positive definite. The quadratic forms $z^\dagger Rz$, $z^\dagger Qz$, and $z^\dagger G_pz$ have direct physical meanings: they give, respectively, the outcome probability for the pure state $|\Psi\rangle$, the corresponding unnormalized
energy above the ground level, and the outcome probability for the noisy state $\rho_{TM}$. Since complex conjugation in the chosen product basis preserves both normalization and product structure, $z$ is a product vector if and only if $c$ is.

For normalized $z$, the success probability, conditional capacity, and conditional ergotropy take the form
\begin{align}
	P_v
	&=z^\dagger G_p z,\\
	\capc(\rho_T^v;H_T)
	&=(1-p)\Delta E\,
	\frac{z^\dagger Rz}{z^\dagger G_p z},\\
	\ee(\rho_T^v;H_T)
	&=(1-p)
	\frac{z^\dagger Qz}{z^\dagger G_p z},
	\label{eq:rayleighObjectives}
\end{align}
where $\Delta E\coloneqq E_{\max}-E_{\min}$.

\begin{proposition}\label{prop:attainability}
	Let $0<p<1$ and $\Delta E>0$, and define
	\begin{align}
		\lambda_p
		&\coloneqq
		\lambda_{\max}\left(G_p^{-1/2}QG_p^{-1/2}\right),\\
		\mathcal V_p
		&\coloneqq
		\ker \left(Q-\lambda_p G_p\right).
		\label{eq:optimalSpace}
	\end{align}
The maximum conditional ergotropy over all joint measurement vectors is $(1-p)\lambda_p$. 
A product measurement outcome attains this value if and only if $\mathcal V_p$ contains a nonzero product vector. 
Otherwise, the product optimum is strictly smaller than the unrestricted joint optimum.
\end{proposition}

The remaining optimization problems reduce to ordinary Rayleigh quotients.
The normalized maximizers of $P_v$, $\mathcal C(\rho_T^v;H_T)$, and $Y_{\mathcal C,v}$ are the unit vectors in the eigenspace of $R$ associated with its largest eigenvalue, whereas the maximizers of $Y_{\mathcal E,v}$ are the unit vectors in the maximal eigenspace of $Q$.

The product attainability criterion for conditional ergotropy becomes particularly simple when the auxiliary system consists of two qubits. 
Every complex subspace of dimension at least two contains a product vector \cite{Parthasarathy2004}.
Hence, a strict product joint gap occurs if and only if $\mathcal V_p$ is one-dimensional and spanned by an entangled vector.
 Writing a generator of $\mathcal V_p$ as
\begin{align}
	z=(z_{00},z_{01},z_{10},z_{11})^{\mathsf T},
\end{align}
this condition is equivalent to
\begin{align}
	z_{00}z_{11}-z_{01}z_{10}\neq0.
\end{align}

The product condition must therefore be checked directly on $z$.
Although the generalized eigenvalue problem can be converted into an ordinary eigenvalue problem by the change of variables $y\coloneqq G_p^{1/2}z$, this transformation does not in general preserve product structure. 
The proof is given in Appendix~\ref{app:optimization}.~Sec.~\ref{app:criterion}, using the generalized Rayleigh principle together with the determinant criterion for two qubit product vectors. 
A common optimizer for the different quantities need not exist in general, although all examples considered below admit one under unrestricted joint optimization. 
The endpoints $p=0$ and $p=1$ require separate treatment.


\subsection{Averages over complete measurements}

For the global white noise family in Eq.~\eqref{eq:generalNoise}, the outcome averaged capacity and ergotropy are independent of the basis chosen for a complete projective measurement on $M$.

\begin{proposition}\label{prop:averages}
For any state of the form in Eq.~\eqref{eq:generalNoise} and any complete rank-one projective measurement on $M$,
	\begin{align}
		\sum_iP_i\capc(\rho_T^i;H_T)
		&=(1-p)(E_{\max}-E_{\min}),
		\label{eq:averageC}\\
		\sum_iP_i\ee(\rho_T^i;H_T)
		&=(1-p)\bigl[\Tr(\rho_T^{\Psi}H_T)-E_{\min}\bigr].
		\label{eq:averageE}
	\end{align}
	Outcomes of zero probability are understood to contribute zero to the sums.
\end{proposition}

Summing Eqs.~\eqref{eq:generalWeightedC} and \eqref{eq:generalWeightedE} over all outcomes and using
\begin{align}
	\sum_i|\chi_i\rangle\langle\chi_i|
	=
	\rho_T^\Psi
	\label{eq:mainCompleteness}
\end{align}
immediately gives both identities. A complementary spectral derivation is provided in Appendix~\ref{app:optimization}.~Sec.~\ref{app:averages}. At $p=0$, the ergotropy relation reduces to the known result for pure resource states \cite{Francica2017,Bernards2019}. Related measurement basis independence of complete outcome averages has also appeared in auxiliary assisted energy distillation protocols \cite{Chaki2025,Chaki2026}. Here, Prop.~\ref{prop:averages} establishes exact basis independent identities for both the outcome averaged capacity and ergotropy throughout the global white noise family. Any advantage offered by joint measurements in these energetic quantities must therefore arise from postselection.


\section{Single qubit batteries from noisy three qubit resources}\label{sec:three}

We first consider qubit $A$ as the battery and measure the auxiliary pair $BC$.
For any $0<p<1$, the noisy $\mathrm{W}_3$ and GHZ$_3$ states exhibit qualitatively different behavior: the optimal conditional ergotropy requires an entangled measurement outcome for $\mathrm{W}_3$, whereas a product outcome is sufficient for GHZ$_3$.

The battery Hamiltonian is
\begin{align}
	H_A\coloneqq\epsilon_A\sigma_z,\quad \epsilon_A>0,
	\label{eq:HA}
\end{align}
where $\sigma_z\coloneqq|0\rangle\langle0|-|1\rangle\langle1|$. 
Thus, the two energy levels are separated by $2\epsilon_A$, with $|0\rangle$ and $|1\rangle$ denoting the upper and lower levels, respectively.


\subsection{Noisy W state: a strict gap between the product and joint optima}\label{subsec:W3}

For the W$_3$ state, the auxiliary states $|01\rangle_{BC}$ and $|10\rangle_{BC}$ are both correlated with the upper battery level $|0\rangle_A$.
A joint measurement on $BC$ can therefore access coherent superpositions of these two components, which is not generally possible with product measurement outcomes.
We consider
\begin{align}
	\rho_p
	&\coloneqq(1-p)|\mathrm{W}_3\rangle\langle\mathrm{W}_3|+\frac{p}{8}\1_8,\\
	|\mathrm{W}_3\rangle
	&\coloneqq\frac{|100\rangle+|010\rangle+|001\rangle}{\sqrt{3}}.
	\label{eq:W3}
\end{align}
Here and throughout this section, $0\leqslant p\leqslant1$. 
Before any measurement on $BC$, the reduced battery state has eigenvalues $(2+p)/6$ and $(4-p)/6$. 
Its capacity and ergotropy therefore coincide,
\begin{align}
	\capc(\rho_p^A;H_A)
	=
	\ee(\rho_p^A;H_A)
	=
	\frac{2(1-p)}{3}\epsilon_A.
	\label{eq:W3before}
\end{align}

\noindent\textbf{Conditional  branch.}
Let
$s\coloneqq c_{21}+c_{31}$,
which collects the two amplitudes in the odd parity subspace of $BC$. 
We then define
\begin{align}
	N
	\coloneqq
	|c_{11}|^2+|s|^2
	=
	1+2\operatorname{Re}(c_{21}\overline{c_{31}})-|c_{41}|^2,
	\label{eq:WN}
\end{align}
After projection onto $|v\rangle$, the pure component of the battery branch has upper and lower level weights $|s|^2/3$ and $|c_{11}|^2/3$, respectively. The corresponding unnormalized conditional state is
\begin{align}
	P_v\rho_A^v
	=
	\frac{p}{8}\1_2
	+
	\frac{1-p}{3}|w_v\rangle\langle w_v|,
	\label{eq:Wconditional}
\end{align}
where
$|w_v\rangle
\coloneqq
\overline{(c_{21}+c_{31})}|0\rangle_A
+
\overline{c_{11}}|1\rangle_A$
is the unnormalized pure component on battery $A$, with
$N=\langle w_v|w_v\rangle$.
It follows that
\begin{align}
	P_v
	&=
	\frac{p}{4}
	+
	\frac{1-p}{3}N,
	\label{eq:WP}\\
	\capc(\rho_A^v;H_A)
	&=
	\frac{8(1-p)N}
	{3p+4(1-p)N}\epsilon_A,
	\label{eq:WC}\\
	\ee(\rho_A^v;H_A)
	&=
	\frac{8(1-p)|s|^2}
	{3p+4(1-p)N}\epsilon_A.
	\label{eq:WE}
\end{align}

\noindent\textbf{Optimal measurements.}
The allowed range of $N$ is $0\leqslant N\leqslant2$. 
The upper bound is attained by the Bell vector
\begin{align}
	|v_W^\star\rangle\coloneqq\frac{|01\rangle+|10\rangle}{\sqrt{2}},
	\label{eq:W3optvector}
\end{align}
for which the two amplitudes in the odd parity subspace add constructively. 
This outcome gives
\begin{align}
	P_{\max}
	&=\frac{8-5p}{12},\\
	\capc_{\max}
	&=\ee_{\max}
	=\frac{16(1-p)}{8-5p}\epsilon_A.
	\label{eq:W3max}
\end{align}
The maximal eigenspace of $R$ is one-dimensional and spanned by $|v_W^\star\rangle$. 
For $0<p<1$, the same vector also spans $\mathcal V_p$. 
Since $|v_W^\star\rangle$ is entangled, Prop.~\ref{prop:attainability} implies a strict product joint gap in the conditional ergotropy.

For product measurement outcomes, one has $N\leqslant4/3$. 
The corresponding individual optima are
\begin{align}
	P_{\max}^{\mathrm{prod}}
	&=\frac{16-7p}{36},
	\label{eq:W3productP}\\
	\capc_{\max}^{\mathrm{prod}}
	&=\frac{32(1-p)}{16-7p}\epsilon_A,
	\label{eq:W3productC}\\
	\ee_{\max}^{\mathrm{prod}}
	&=\frac{8(1-p)}{4-p}\epsilon_A.
	\label{eq:W3product}
\end{align}
The probability and capacity optima are attained by $(\sqrt{2/3}|0\rangle+|1\rangle/\sqrt3)^{\otimes2}$,
whereas the ergotropy optimum is attained by either $|01\rangle$ or $|10\rangle$. All of these product outcomes are already available within Protocol I.
Protocols I and II therefore give the same individual optima, although no single product vector simultaneously attains all three for $0<p<1$.

For $0<p<1$, the unrestricted joint optimum is strictly larger than the product optimum for each of the three quantities. 
At $p=0$, the conditional capacity and ergotropy can also attain their joint optima with product outcomes, while the maximum success probability still requires the Bell outcome.
At the joint optimum,
\begin{align}
	Y_{\capc,\max}
	=
	Y_{\ee,\max}
	=
	\frac{4(1-p)}{3}\epsilon_A.
	\label{eq:W3yield}
\end{align}
The selected Bell branch carries the entire outcome-averaged ergotropy in Prop.~\ref{prop:averages}, whereas the averaged capacity also receives contributions from the remaining outcomes.

\begin{figure*}[t]
\centering
\begin{minipage}[t]{0.475\linewidth}
 \centering
 \includegraphics[width=\linewidth]{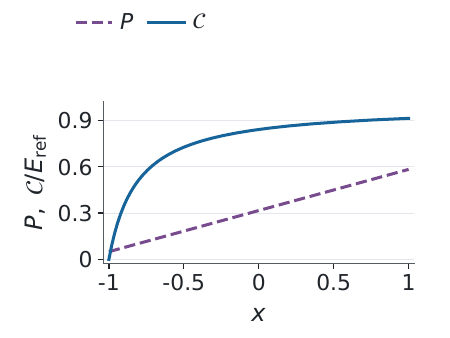}\\[-0.2em]
 \textbf{(a)}
\end{minipage}\hfill%
\begin{minipage}[t]{0.475\linewidth}
 \centering
 \includegraphics[width=\linewidth]{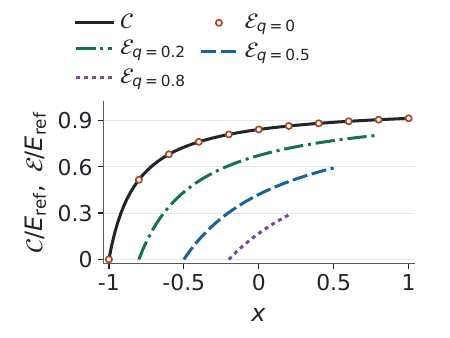}\\[-0.2em]
 \textbf{(b)}
\end{minipage}
\caption{
Single qubit battery obtained from the noisy W$_3$ state for $p=0.2$ and $\epsilon_A=0.5E_{\mathrm{ref}}$. 
Energies are given in units of $E_{\mathrm{ref}}$, while probabilities are dimensionless.
(a) Success probability and conditional capacity as functions of $x=N-1$; the endpoint $x=1$ corresponds to the Bell outcome in Eq.~\eqref{eq:W3optvector}. 
(b) Conditional capacity and ergotropy for fixed $q=|c_{11}|^2$, with the allowed region $q-1\leqslant x\leqslant1-q$. 
Open circles mark the ergotropy values at $q=0$, where the ergotropy coincides with the capacity.
}
\label{fig:W3performance}
\end{figure*}

For product measurement outcomes, $Y_{\capc}\leqslant8(1-p)\epsilon_A/9$ and 
$Y_{\ee}\leqslant2(1-p)\epsilon_A/3$.
Both bounds are strictly below the corresponding joint optimum for $0\leqslant p<1$.
To parametrize the performance landscape in Fig.~\ref{fig:W3performance}, define
\begin{align}
x\coloneqq
2\operatorname{Re}(c_{21}\overline{c_{31}})
	-|c_{41}|^2
=
	N-1,\label{eq:Wx}
\end{align}
with $-1\leqslant x\leqslant1$.
For fixed $q\coloneqq|c_{11}|^2$, normalization together with the Cauchy-Schwarz inequality gives $q-1\leqslant x\leqslant1-q$.
Equation~\eqref{eq:WE} can then be written as
\begin{align}
	\ee(\rho_A^v;H_A)
	=
	\frac{8(1-p)(1+x-q)}
	{3p+4(1-p)(1+x)}
	\epsilon_A.
	\label{eq:WExq}
\end{align}
When $q=0$ and $N>0$, the projected pure component is proportional to $|0\rangle_A$. 
The conditional ergotropy therefore equals the capacity, even though the postselected battery state is generally mixed.

\subsection{Noisy GHZ state: a product measurement outcome suffices}\label{subsec:GHZ3}

For the GHZ$_3$ resource, only the auxiliary basis states $|00\rangle_{BC}$ and $|11\rangle_{BC}$ contribute to the projected pure component on battery $A$. 
We consider
\begin{align}
	\rho_p
	&\coloneqq
	(1-p)|\mathrm{GHZ}_3\rangle\langle\mathrm{GHZ}_3|+\frac{p}{8}\1_8,
	\label{eq:GHZ3}
\end{align}
where
$|\mathrm{GHZ}_3\rangle\coloneqq(|000\rangle+|111\rangle)/\sqrt{2}$,
with $0\leqslant p\leqslant1$.
Before any measurement on $BC$, the reduced state of the battery is maximally mixed. 
Its capacity and ergotropy therefore both vanish.

\noindent\textbf{Conditional quantities.}
Let $m\coloneqq|c_{11}|^2+|c_{41}|^2$ denote the weight of the measurement vector in the even parity subspace of $BC$, with $|c_{11}|^2$ and $|c_{41}|^2$ giving the upper and lower level weights of the projected pure component on battery $A$, respectively.
The unnormalized conditional state is
\begin{align}
	P_v\rho_A^v
	=
	\frac{p}{8}\1_2
	+
	\frac{1-p}{2}|g_v\rangle\langle g_v|,
	\label{eq:GHZ3state}
\end{align}
where
\begin{align}
	|g_v\rangle
	\coloneqq
	\overline{c_{11}}|0\rangle_A
	+
	\overline{c_{41}}|1\rangle_A.
	\label{eq:GHZ3vector}
\end{align}
The vector $|g_v\rangle$ is the unnormalized pure component on battery $A$, with
$\langle g_v|g_v\rangle=m$. 
The projection of $|\mathrm{GHZ}_3\rangle$ onto the selected auxiliary outcome $|v\rangle$ is $|g_v\rangle/\sqrt2$; for $m>0$, the corresponding normalized battery state is $|g_v\rangle/\sqrt m$.

It follows that the success probability, conditional capacity, and conditional ergotropy are
\begin{align}
	P_v
	&=
	\frac{p+2(1-p)m}{4},
	\label{eq:GHZ3P}\\
	\capc(\rho_A^v;H_A)
	&=
	\frac{4(1-p)m}{p+2(1-p)m}\epsilon_A,
	\label{eq:GHZ3C}\\
	\ee(\rho_A^v;H_A)
	&=
	\frac{4(1-p)|c_{11}|^2}{p+2(1-p)m}\epsilon_A.
	\label{eq:GHZ3E}
\end{align}

\noindent\textbf{Optimal measurements.}
All three quantities are simultaneously maximized at $m=|c_{11}|^2=1$, corresponding to the product outcome $|v\rangle=|00\rangle$. 
This gives 
\begin{align}
	P_{\max}
	&=\frac{2-p}{4},\\
	\capc_{\max}
	&=\ee_{\max}
	=\frac{4(1-p)}{2-p}\epsilon_A.
	\label{eq:GHZ3max}
\end{align}
For $0<p<1$, the maximizing generalized eigenspace is
$\mathcal V_p=\operatorname{span}\{|00\rangle\}$. 
Since $|00\rangle$ is a product vector and is allowed in all three measurement classes, no joint measurement advantage arises in this case. 
The same outcome also maximizes the probability weighted capacity and ergotropy, yielding 
$Y_{\capc,\max}=Y_{\ee,\max}=(1-p)\epsilon_A$.
For the capacity, this agrees with the local projective measurement optimum reported in Ref.~\cite{Wang2025}.

\begin{figure*}[t]
\centering
\begin{minipage}[t]{0.475\linewidth}
 \centering
 \includegraphics[width=\linewidth]{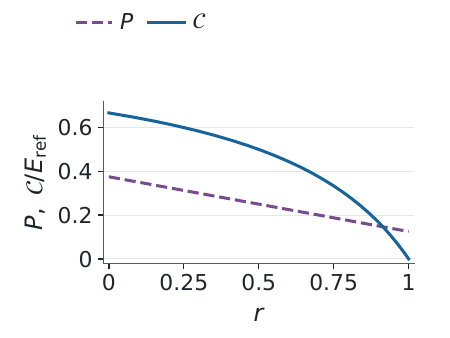}\\[-0.2em]
 \textbf{(a)}
\end{minipage}\hfill%
\begin{minipage}[t]{0.475\linewidth}
 \centering
 \includegraphics[width=\linewidth]{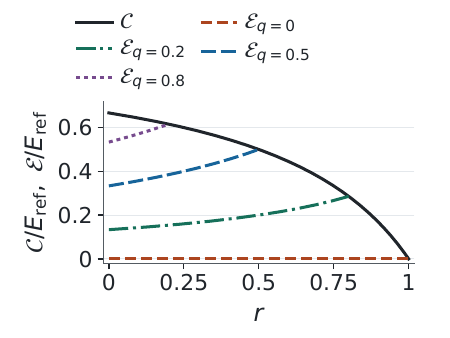}\\[-0.2em]
 \textbf{(b)}
\end{minipage}
\caption{
Single qubit battery obtained from the noisy GHZ$_3$ resource for $p=0.5$ and $\epsilon_A=0.5E_{\mathrm{ref}}$. 
Energies are given in units of $E_{\mathrm{ref}}$, while probabilities are dimensionless. 
(a) Success probability and conditional capacity as functions of the weight $r=1-m$ outside the even parity subspace of the auxiliary pair.
(b) Conditional capacity and ergotropy for fixed $q=|c_{11}|^2$; each ergotropy curve terminates at the physical boundary $r=1-q$.
}
\label{fig:GHZ3performance}
\end{figure*}

For Fig.~\ref{fig:GHZ3performance}, we parametrize the measurement vector by
\begin{align}
	r&\coloneqq|c_{21}|^2+|c_{31}|^2=1-m,\\
	q&\coloneqq|c_{11}|^2.
	\label{eq:GHZrq}
\end{align}
Normalization then gives the allowed range
$0\leqslant r\leqslant1-q$.
For $0<p<1$ and fixed $q>0$, increasing $r$ decreases both the success probability and the conditional capacity, while increasing the conditional ergotropy. 
The constraint $r\leqslant 1-q$, however, ensures that this local increase in ergotropy does not alter the global optimum.

\section{Two qubit batteries from noisy four qubit resources}\label{sec:four}

We next consider $AB$ as an interacting two qubit battery and measure the auxiliary pair $CD$. 
In contrast to the single qubit case, the interaction between the battery qubits can change which postselected outcome maximizes the conditional ergotropy. 
This dependence arises because ergotropy is sensitive not only to the spectrum of the conditional state but also to the energy expectation of its projected pure component.

The battery Hamiltonian is
\begin{align}
	H_{AB}
	\coloneqq
	\begin{aligned}[t]
		\epsilon_A\sigma_z\otimes\1_2
		+\epsilon_B\1_2\otimes\sigma_z
	    +J\sigma_x\otimes\sigma_x
	\end{aligned}
	\label{eq:HAB}
\end{align}
where $\sigma_x\coloneqq|0\rangle\langle1|+|1\rangle\langle0|$. 
The parameters $\epsilon_A$ and $\epsilon_B$ are half the local level spacings, while $J$ denotes the interaction strength between the two battery qubits. 
Throughout this section, we take $\epsilon_A,\epsilon_B,J\geqslant0$.

Excluding the trivial case of a vanishing Hamiltonian, let $S\coloneqq\epsilon_A+\epsilon_B$ and $D\coloneqq\epsilon_A-\epsilon_B$, and define
\begin{align}
	\Omega_+
	&\coloneqq
	\sqrt{S^2+J^2},
	&
	\Omega_-
	&\coloneqq
	\sqrt{D^2+J^2}.
	\label{eq:omegas}
\end{align}
Since $\Omega_+\geqslant\Omega_-$, the energy spectrum, in increasing order, is $-\Omega_+,-\Omega_-,\Omega_-,\Omega_+$. The even and odd parity sectors have eigenvalues $\pm\Omega_+$ and $\pm\Omega_-$, respectively. 
In what follows, the coefficients $c_{k1}$ refer to the measurement vector on the auxiliary pair $CD$.


\subsection{Noisy GHZ state: interaction induced measurement entanglement}\label{subsec:GHZ4}

For the GHZ$_4$ state, postselection prepares a pure component entirely within the even parity sector of the battery. 
The interaction couples $|00\rangle$ and $|11\rangle$, so the coherence between these two components directly affects the conditional ergotropy. 
We consider
\begin{align}
	\rho_p
	&\coloneqq
	(1-p)|\mathrm{GHZ}_4\rangle\langle\mathrm{GHZ}_4|
	+\frac{p}{16}\1_{16},
	\label{eq:GHZ4}
\end{align}
where
$|\mathrm{GHZ}_4\rangle\coloneqq(|0000\rangle+|1111\rangle)/\sqrt{2}$.
Before any measurement on the auxiliary pair, the reduced battery state has two doubly degenerate eigenvalues, $(2-p)/4$ and $p/4$. 
Its capacity and ergotropy are therefore
\begin{align}
	\capc(\rho_p^{AB};H_{AB})
	&=(1-p)(\Omega_++\Omega_-),\\
	\ee(\rho_p^{AB};H_{AB})
	&=\frac{1-p}{2}(\Omega_++\Omega_-).
	\label{eq:GHZ4before}
\end{align}

\noindent\textbf{Conditional branch.}
With $m\coloneqq|c_{11}|^2+|c_{41}|^2$, the unnormalized conditional state is
\begin{align}
	P_v\rho_{AB}^v
	&=
	\frac{p}{16}\1_4
	+\frac{1-p}{2}|G_v\rangle\langle G_v|,
	\label{eq:GHZ4state}
\end{align}
where
$|G_v\rangle\coloneqq\overline{c_{11}}|00\rangle_{AB}
+\overline{c_{41}}|11\rangle_{AB}$.
The vector $|G_v\rangle$ is the unnormalized pure component on the battery, with
$\langle G_v|G_v\rangle=m$. 
The projection of $|\mathrm{GHZ}_4\rangle$ onto the selected auxiliary outcome $|v\rangle$ is $|G_v\rangle/\sqrt2$; for $m>0$, the corresponding normalized battery state is $|G_v\rangle/\sqrt m$.

For $m>0$, define
$t\coloneqq|c_{11}|^2/m$ and
$\phi\coloneqq\arg(c_{11})-\arg(c_{41})$.
Here, $m$ is the weight of the measurement vector in the even parity subspace, while $t$ gives the population of $|00\rangle_{AB}$ in the normalized projected state. 
The relative phase $\phi$ controls the coherence between $|00\rangle_{AB}$ and $|11\rangle_{AB}$; it is irrelevant at $t=0$ or $1$. 
When $m=0$ and $p>0$, the conditional battery state is maximally mixed.

Writing $\ee_v\coloneqq\ee(\rho_{AB}^v;H_{AB})$, the success probability, conditional capacity, and conditional ergotropy are
\begin{align}
	&P_v
	=
	\frac{p+2(1-p)m}{4},
	\label{eq:GHZ4P}\\
    &\capc(\rho_{AB}^v;H_{AB})
	=
	\frac{4(1-p)m\Omega_+}
	{p+2(1-p)m},
	\label{eq:GHZ4C}\\
    &\ee(\rho_{AB}^v;H_{AB})
	=
	\tfrac{2(1-p)m}{p+2(1-p)m}\\
	&\qquad\cdot
	\Bigl[
	(2t-1)S+\Omega_+
	+2\sqrt{t(1-t)}J\cos\phi
	\Bigr].
	\label{eq:GHZ4E}
\end{align}

\noindent\textbf{Optimal measurements.}
For fixed $t$ and $\phi$, all three quantities are nondecreasing functions of $m$, so a common optimizer can be chosen with $m=1$. 
The success probability and capacity are then already maximized. 
For $\Omega_+>0$, maximizing the ergotropy further gives $\cos\phi=1$ and
\begin{align}
	t
	&=
	t_\star
	\coloneqq
	\frac{1}{2}
	\left(
	1+\frac{S}{\Omega_+}
	\right).
	\label{eq:GHZ4t}
\end{align}
When $J=0$, the relative phase is irrelevant. 
The joint optima are therefore
\begin{align}
	P_{\max}
	&=
	\frac{2-p}{4},\\
	\capc_{\max}
	&=
	\ee_{\max}
	=
	\frac{4(1-p)\Omega_+}{2-p}.
	\label{eq:GHZ4max}
\end{align}

For $p<1$ and $\Omega_+>0$, the optimal conditional ergotropy is strictly larger than its premeasurement value in Eq.~\eqref{eq:GHZ4before}. 
The conditional capacity is also strictly enhanced, except when $p=0$ and $\Omega_+=\Omega_-$, for which the two capacities coincide.

For $J>0$, the optimal joint outcome is
\begin{align}
	|v_{\mathrm{GHZ}}^\star\rangle
	&\coloneqq
	\sqrt{t_\star}|00\rangle
	+
	\sqrt{1-t_\star}|11\rangle,
	\label{eq:GHZ4vector}
\end{align}
which is entangled and simultaneously attains all three maxima. For $0<p<1$, reaching the maximal conditional ergotropy requires $m=1$ and a measurement vector proportional to Eq.~\eqref{eq:GHZ4vector}; no product vector satisfies both conditions when $J>0$ (Appendix~\ref{app:optimization}). 
By contrast, the product outcome $|00\rangle$ still maximizes both $P_v$ and $\capc(\rho_{AB}^v;H_{AB})$.

The need for an entangled outcome is therefore specific to ergotropy. 
The capacity depends only on the spectrum of the conditional state and is already optimized by a product outcome, whereas the ergotropy is also sensitive to the orientation of the projected state relative to the energy eigenbasis. 
Indeed, the vector in Eq.~\eqref{eq:GHZ4vector} projects the pure component onto the highest energy eigenstate of the even parity sector.

For comparison with the joint optimum, define $p_c\coloneqq2J/(J+\Omega_++S)$.
The exact product optimum for the conditional ergotropy is
\begin{align}
	\ee_{\max}^{\mathrm{prod}}
	&=
	\begin{cases}
		(2-p)\Omega_+-pJ,
		&0\leqslant p<p_c,\\[2pt]
		\displaystyle\frac{2(1-p)(\Omega_++S)}{2-p},
		&p_c\leqslant p\leqslant1.
	\end{cases}
	\label{eq:GHZ4productE}
\end{align}
The two branches meet continuously at $p_c$. 
Below this threshold, the optimum is attained by a tilted product vector, whereas for $p\geqslant p_c$, the product outcome $|00\rangle$ is optimal.

For the success probability weighted ergotropy, the product and joint optima are
\begin{align}
	Y_{\ee,\max}^{\mathrm{prod}}
	&=
	\frac{1-p}{2}(\Omega_++S),\\
	Y^{\mathrm{joint}}_{\ee,\max}
	&=
	(1-p)\Omega_+,
	\label{eq:GHZ4productYield}
\end{align}
with the product optimum attained by $|00\rangle$. 
Thus, maximizing the conditional ergotropy and maximizing its probability weighted contribution can select different product measurement outcomes.

For $J>0$ and $0<p<1$, the product and joint optima of the conditional ergotropy are strictly separated. 
The corresponding weighted optima are also strictly separated for $J>0$ and $0\leqslant p<1$. 
At $p=0$ and $J>0$, a product outcome can attain the maximal conditional ergotropy, but only with a smaller success probability than the joint optimum. 
At $p=1$, all energetic quantities vanish. 
Appendix~\ref{app:optimization}.~Sec.~\ref{app:GHZ4product} gives the full derivation of the product bound, including the limiting cases $J=0$ and $S=0$.

\begin{figure*}[t]
\centering
\begin{minipage}[t]{0.475\linewidth}
 \centering
 \includegraphics[width=\linewidth]{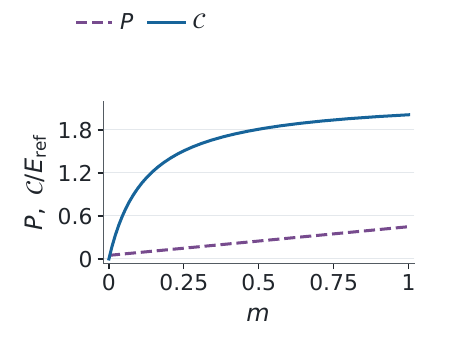}\\[-0.2em]
 \textbf{(a)}
\end{minipage}\hfill%
\begin{minipage}[t]{0.475\linewidth}
 \centering
 \includegraphics[width=\linewidth]{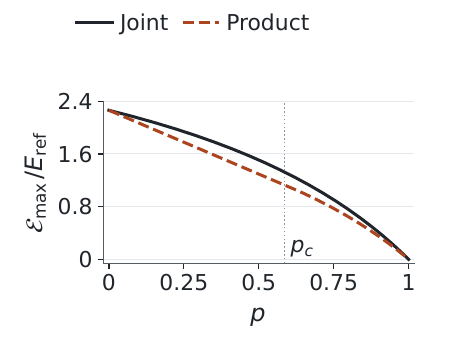}\\[-0.2em]
 \textbf{(b)}
\end{minipage}
\caption{
Two qubit battery obtained from the noisy GHZ$_4$ resource for $\epsilon_A=0.5E_{\mathrm{ref}}$, $\epsilon_B=0.3E_{\mathrm{ref}}$, and $J=0.8E_{\mathrm{ref}}$. 
Energies are given in units of $E_{\mathrm{ref}}$, while probabilities are dimensionless.
(a) Success probability and conditional capacity as functions of the weight $m$ in the even parity subspace for $p=0.2$. 
(b) Maximum conditional ergotropy over unrestricted joint and product measurement outcomes as a function of the noise strength $p$.
The optimal product outcome changes at $p_c=2J/(J+\Omega_++S)$. 
The joint and product optima coincide at $p=0$ and $p=1$, but are strictly separated for $0<p<1$.
}
\label{fig:GHZ4performance}
\end{figure*}

\subsection{Noisy W state: optimal measurement selected by the interaction}\label{subsec:W4}

For the W$_4$ state, the projected pure component on the battery is supported on the two-dimensional subspace spanned by $|00\rangle_{AB}$ and $|\psi_+\rangle_{AB}\coloneqq(|01\rangle+|10\rangle)/\sqrt2$. 
Their energy expectations are $\langle00|H_{AB}|00\rangle=S$ and $\langle\psi_+|H_{AB}|\psi_+\rangle=J$.
The optimal postselected outcome is therefore determined by the competition between the local energy scale $S$ and the interaction strength $J$.

We consider
\begin{align}
	\rho_p
	&\coloneqq
	(1-p)|\mathrm{W}_4\rangle\langle\mathrm{W}_4|
	+\frac{p}{16}\1_{16},\\
	|\mathrm{W}_4\rangle
	&\coloneqq
	\frac{1}{2}
	\bigl(
	|1000\rangle+|0100\rangle+|0010\rangle+|0001\rangle
	\bigr).
	\label{eq:W4}
\end{align}
Before any measurement on the auxiliary pair, the reduced battery state has two doubly degenerate eigenvalues, $(2-p)/4$ and $p/4$. 
Its capacity and ergotropy are therefore
\begin{align}
	\capc(\rho_p^{AB};H_{AB})
	&=
	(1-p)(\Omega_++\Omega_-),
	\label{eq:W4beforeC}\\
	\ee(\rho_p^{AB};H_{AB})
	&=
	\frac{1-p}{2}
	\bigl(
	S+J+\Omega_++\Omega_-
	\bigr).
	\label{eq:W4beforeE}
\end{align}

\noindent\textbf{Conditional branch.}
Let $u\coloneqq2|c_{11}|^2$, and define
\begin{align}
	m
	\coloneqq
	u+|c_{21}+c_{31}|^2.
	\label{eq:W4mu}
\end{align}
For the pure W$_4$ state, the selected measurement outcome occurs with probability $m/4$.
The corresponding unnormalized conditional state is
\begin{align}
	P_v\rho_{AB}^v
	&=
	\frac{p}{16}\1_4
	+
	\frac{1-p}{4}|W_v\rangle\langle W_v|,
	\label{eq:W4state}
\end{align}
where
\begin{align}
	|W_v\rangle
	&\coloneqq
	\overline{(c_{21}+c_{31})}|00\rangle_{AB}
	+
	\overline{c_{11}}
	\bigl(|01\rangle_{AB}+|10\rangle_{AB}\bigr).
	\label{eq:W4vector}
\end{align}
Here, $|W_v\rangle$ is the unnormalized pure component on battery $AB$, with
$m=\langle W_v|W_v\rangle$. 
The quantities $u$ and $m-u$ are the squared norms of its odd and even parity components, respectively. 
Their allowed domain is $0\leqslant u\leqslant m\leqslant2$.

The success probability, conditional capacity, and conditional ergotropy are therefore
\begin{align}
	&P_v
	=
	\frac{p+(1-p)m}{4},
	\label{eq:W4P}\\
	&\capc(\rho_{AB}^v;H_{AB})
	=
	\frac{2(1-p)m\Omega_+}
	{p+(1-p)m},
	\label{eq:W4C}\\
	&\ee(\rho_{AB}^v;H_{AB})
	=
	\frac{(1-p)
		\left[
		m(S+\Omega_+)+u(J-S)
		\right]}
	{p+(1-p)m}.
	\label{eq:W4E}
\end{align}

\noindent\textbf{Optimal measurements.}
Both $P_v$ and $\capc(\rho_{AB}^v;H_{AB})$ are nondecreasing functions of $m$ and reach their maxima at $m=2$:
\begin{align}
	P_{\max}
	&=
	\frac{2-p}{4},\\
	\capc_{\max}
	&=
	\frac{4(1-p)\Omega_+}{2-p}.
	\label{eq:W4PCmax}
\end{align}
At fixed $m$, the conditional ergotropy is affine in $u$, so the optimizer is determined by the sign of $J-S$. 
For $J<S$, the maximum is attained at $(u,m)=(0,2)$, corresponding to the Bell outcome, whereas for $J>S$ it is attained at $(u,m)=(2,2)$, corresponding to the product outcome $|00\rangle$. 
At the transition point $J=S$, every state on the boundary $m=2$ is optimal.
Hence,
\begin{align}
	\ee_{\max}
	&=
	\begin{cases}
		\displaystyle\frac{2(1-p)(S+\Omega_+)}{2-p},
		&J\leqslant S,\\[0.8em]
		\displaystyle\frac{2(1-p)(J+\Omega_+)}{2-p},
		&J\geqslant S.
	\end{cases}
	\label{eq:W4Emax}
\end{align}

For $0<p<1$, the maximizing generalized eigenspace $\mathcal V_p$ is spanned by the Bell vector when $J<S$ and by $|00\rangle$ when $J>S$.
At $J=S$, $\mathcal V_p$ is the two-dimensional subspace spanned by these two vectors. 
A product measurement outcome can therefore attain the maximum conditional ergotropy if and only if $J\geqslant S$. 
At $p=0$, the product outcome $|01\rangle$ reaches the same conditional capacity and ergotropy as the Bell outcome, but with a smaller success probability.
For $J\geqslant S$, $|00\rangle$ is optimal.

The maximal conditional capacity and ergotropy are no smaller than their premeasurement values in Eqs.~\eqref{eq:W4beforeC} and \eqref{eq:W4beforeE}. 
The maximum probability-weighted ergotropy is $Y_{\ee,\max}=(1-p)[\Omega_++\max(S,J)]/2$.
The measurement transition at $J=S$ therefore reflects a crossing between the energy expectations of the two accessible battery components, $|00\rangle$ and $|\psi_+\rangle$.

\begin{figure*}[t]
\centering
\begin{minipage}[t]{0.475\linewidth}
 \centering
 \includegraphics[width=\linewidth]{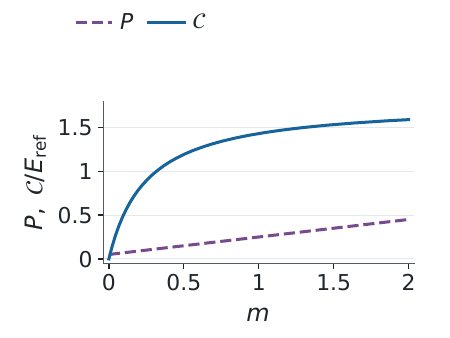}\\[-0.2em]
 \textbf{(a)}
\end{minipage}\hfill%
\begin{minipage}[t]{0.475\linewidth}
 \centering
 \includegraphics[width=\linewidth]{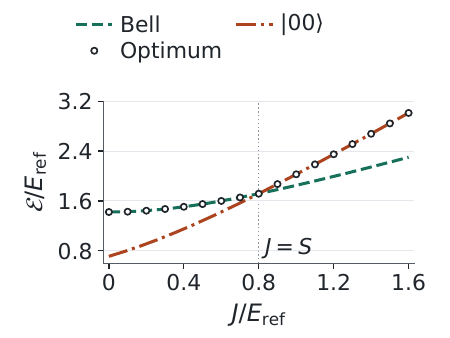}\\[-0.2em]
 \textbf{(b)}
\end{minipage}
\caption{
Two qubit battery obtained from the noisy W$_4$ resource for $p=0.2$, $\epsilon_A=0.5E_{\mathrm{ref}}$, and $\epsilon_B=0.3E_{\mathrm{ref}}$. 
Energies are given in units of $E_{\mathrm{ref}}$, while probabilities are dimensionless. 
(a) Success probability and conditional capacity as functions of $m$ for $J=0.4E_{\mathrm{ref}}$.
(b) Conditional ergotropy at $m=2$ as a function of the interaction strength $J$. 
Open circles mark the unrestricted joint optimum, given by the upper envelope of the Bell branch ($u=0$) and the $|00\rangle$ branch ($u=2$). 
The vertical dotted line indicates the transition point $J=S=0.8E_{\mathrm{ref}}$.
}
\label{fig:W4performance}
\end{figure*}

In Fig.~\ref{fig:W4performance}(b), the upper envelope of the Bell and $|00\rangle$ branches gives the unrestricted joint optimum. 
For $J<S$, however, the displayed $|00\rangle$ branch need not coincide with the optimum over all product measurement outcomes.

Table~\ref{tab:optima} summarizes a representative measurement outcome that simultaneously maximizes the success probability, conditional capacity, and conditional ergotropy in each case. 
Simultaneous attainability is more restrictive than optimizing either conditional quantity separately. 
In particular, at zero noise, a product outcome can sometimes attain the optimal conditional capacity or ergotropy while having a smaller success probability and therefore failing to maximize all three quantities at once.

\begin{table*}[t]
	\caption{Representative measurement outcomes that simultaneously maximize the success probability, conditional capacity, and conditional ergotropy. 
	The last column indicates whether all three unrestricted joint optima can be attained by a product measurement outcome.}
	\label{tab:optima}
	\begin{tabular}{@{}p{0.14\linewidth}p{0.42\linewidth}p{0.36\linewidth}@{}}
		\toprule
		State & One simultaneous optimizer & Can a product outcome attain all three joint optima?\\
		\midrule
		$\mathrm{W}_3$
		& $(|01\rangle+|10\rangle)/\sqrt{2}$
		& No for $0\leqslant p<1$\\
		
		$\mathrm{GHZ}_3$
		& $|00\rangle$
		& Yes\\
		
		$\mathrm{GHZ}_4$
		& Eq.~\eqref{eq:GHZ4vector}
		& Yes only if $J=0$ or $p=1$\\
		
		$\mathrm{W}_4$, $J<S$
		& $(|01\rangle+|10\rangle)/\sqrt{2}$
		& No for $0\leqslant p<1$\\
		
		$\mathrm{W}_4$, $J=S$
		& Bell or $|00\rangle$
		& Yes\\
		
		$\mathrm{W}_4$, $J>S$
		& $|00\rangle$
		& Yes\\
		\bottomrule
	\end{tabular}
\end{table*}

\section{Conclusions}\label{sec:conclusion}

We have established when product postselection on the auxiliary system can attain the unrestricted joint optimum of conditional ergotropy. 
For a pure battery auxiliary state admixed with global white noise and $0<p<1$, the optimization reduces to a generalized eigenvalue problem. 
The joint optimum is attainable by a product outcome if and only if the eigenspace associated with the largest generalized eigenvalue contains a product vector. 
For two auxiliary qubits, this condition becomes especially simple: a strict product joint gap occurs exactly when the maximizing eigenspace is one-dimensional and spanned by an entangled vector.

The examples show that whether entangled postselection is necessary depends on both the resource state and the battery Hamiltonian. 
For W$_3$, an entangled outcome is required throughout $0<p<1$, whereas GHZ$_3$ admits a product optimum. For interacting two qubit batteries, GHZ$_4$ requires an entangled optimal outcome whenever $J>0$ and $0<p<1$, while for W$_4$ the optimum switches from a Bell outcome to $|00\rangle$ as $J$ crosses $\epsilon_A+\epsilon_B$. 
These examples also highlight the different roles of capacity and ergotropy: capacity depends only on the spectrum of the conditional state, whereas ergotropy is also sensitive to its orientation relative to the energy eigenbasis. 
As a result, the two quantities can favor different measurement outcomes.
Even within the product class, the outcome that maximizes the conditional ergotropy need not maximize its success probability weighted contribution.

For the global white noise family, the capacity and ergotropy averaged over all outcomes of a complete projective measurement are independent of the measurement basis. 
Joint measurements therefore cannot increase these averages; instead, they shift them across the postselected branches and can concentrate a larger share of the energetic performance into a chosen outcome.
In this sense, the joint measurement advantage identified here is inherently postselective.
It will be important to determine how this picture changes for more general noise models, higher rank or nonprojective measurements, and protocols that explicitly account for the energetic costs of measurement and reset.

\begin{acknowledgments}
This work was supported by the Research Fund of Jianghan University (No. 2023JCYJ08).

\end{acknowledgments}

\section*{Author contributions}
Y.D. and Z.Z. performed the analytical calculations and numerical verification.
X.H. conceived and supervised the project. 
All authors discussed the results and contributed to writing and revising the manuscript.

\section*{Conflict of interest}
The authors declare no conflict of interest.

\section*{Data availability statement}
No external datasets were used in this study.
All results are derived analytically, and the data underlying the figures can be reproduced from the expressions provided in the manuscript.

\section*{Use of generative AI}
During the preparation of this manuscript, the authors used ChatGPT (OpenAI) for language editing and LaTeX formatting. All generated text was reviewed and revised by the authors, who take full responsibility for the content of the manuscript.

\clearpage
\onecolumngrid
\appendix
\makeatletter
\renewcommand{\p@subsection}{}
\renewcommand{\p@subsubsection}{\thesubsection~}
\makeatother
\section{}
\label{app:optimization}

This appendix provides the derivations underlying the results in the main text. 
We first establish the spectral identities and the basis independent averages over complete projective measurements, and then prove the criterion for product attainability of the unrestricted joint optimum. 
We subsequently derive the conditional quantities and constrained optima for the four resource states considered in the main text, including the relevant boundary cases.

\subsection{Spectral identities and averages over all outcomes}
\label{app:averages}

All four examples share the same conditional spectral structure.
For a selected outcome $v$ with $P_v>0$, let $\rho_v$ denote the normalized conditional state of a $d$-dimensional battery and write its unnormalized state as
\begin{align}
	\sigma_v
	\coloneqq
	P_v\rho_v
	=
	\eta\1_d
	+
	\kappa|\psi_v\rangle\langle\psi_v|.
	\label{eq:appRankOne}
\end{align}
Here, $|\psi_v\rangle$ is the unnormalized pure component of the selected battery branch, with squared norm $n_v	\coloneqq
\langle\psi_v|\psi_v\rangle$, while $\eta,\kappa\geqslant0$ set the weights of the isotropic background and the rank-one contribution, respectively. 
For Eq.~\eqref{eq:generalBranch}, $\eta=p/(d_Td_M)$, $\kappa=1-p$, and
$|\psi_v\rangle=|\chi_v\rangle$.
 In the examples below, different conventions are used for the normalization of $|\psi_v\rangle$; the corresponding factors are absorbed into $\kappa$.

Taking the trace of Eq.~\eqref{eq:appRankOne} gives
\begin{align}
	P_v
	=
	d\eta+\kappa n_v.
	\label{eq:appRankOneP}
\end{align}
For $P_v>0$ and $n_v>0$, the normalized conditional state has one eigenvalue
\begin{align}
	\lambda_+
	=
	\frac{\eta+\kappa n_v}{P_v}
\end{align}
along the direction of $|\psi_v\rangle$, while the orthogonal subspace carries the $(d-1)$-fold degenerate eigenvalue
\begin{align}
	\lambda_-
	=
	\frac{\eta}{P_v}.
\end{align}
Their separation is therefore
\begin{align}
	\Delta_v
	\coloneqq
	\lambda_+-\lambda_-
	=
	\frac{\kappa n_v}{P_v}.
	\label{eq:appDelta}
\end{align}
If $n_v=0$ and $\eta>0$, the conditional state is maximally mixed and $\Delta_v=0$. 
If $\eta=n_v=0$, the outcome occurs with zero probability, so no normalized conditional state is defined.

For an arbitrary battery Hamiltonian $H$ with extremal eigenvalues $E_{\min}$ and $E_{\max}$, the isotropic background contributes the same term $\eta\Tr H$ to every unitary rearrangement of the spectrum. 
The only state dependent contribution comes from the excess eigenvalue weight $\kappa n_v$, which is placed on the lowest energy level in the passive state and on the highest energy level in the active state.
Hence,
\begin{align}
	P_v E_{\rm passive}
	&=
	\eta\Tr H
	+
	\kappa n_v E_{\min},\\[4pt]
	P_v E_{\rm active}
	&=
	\eta\Tr H
	+
	\kappa n_v E_{\max},\\[4pt]
	P_v\Tr(\rho_vH)
	&=
	\eta\Tr H
	+
	\kappa\langle\psi_v|H|\psi_v\rangle.
\end{align}
Here, $E_{\rm passive}$ and $E_{\rm active}$ denote the passive and active energies of the normalized conditional state $\rho_v$. 
Subtracting the passive energy from the active and actual energies gives Eqs.~\eqref{eq:generalWeightedC} and \eqref{eq:generalWeightedE}, respectively. 
These identities remain valid for Hamiltonians with nonzero trace, since the isotropic contribution $\eta\Tr H$ cancels in the relevant energy differences.

For a complete projective measurement on the auxiliary subsystem, completeness of the measurement basis implies
\begin{align}
	\sum_i|\chi_i\rangle\langle\chi_i|
	=
	\Tr_M \left[
	\left(
	\1_T\otimes\sum_i|v_i\rangle\langle v_i|
	\right)
	|\Psi\rangle\langle\Psi|\right]
	=
	\rho_T^{\Psi}.
\end{align}
Taking the trace gives $\sum_i n_i=1$, which immediately yields Eq.~\eqref{eq:averageC}. 
Likewise, multiplying the completeness relation by $H_T-E_{\min}\1_T$ and taking the trace gives Eq.~\eqref{eq:averageE}. 
If an outcome has zero probability, its unnormalized conditional operator vanishes, so the weighted identities remain well defined at $p=0$. 
At $p=1$, both sides vanish identically. This completes the proof of Prop.~\ref{prop:averages}. 
The result relies on the specific structure of a pure state admixed with global white noise and does not hold for a general mixed resource state.

For the qubit Hamiltonian in Eq.~\eqref{eq:HA}, with ordered energies $-\epsilon_A$ and $+\epsilon_A$, Eqs.~\eqref{eq:capacity} and \eqref{eq:ergotropy} reduce to
\begin{align}
	\capc(\rho_v;H_A)
	&=
	2\epsilon_A\Delta_v,
	\label{eq:appQubitC}\\
	\ee(\rho_v;H_A)
	&=
	\Delta_v
	\left(
	\frac{\langle\psi_v|H_A|\psi_v\rangle}{n_v}
	+\epsilon_A
	\right).
	\label{eq:appQubitE}
\end{align}
When the pure contribution vanishes, the second expression is understood by continuity whenever $P_v>0$.
Writing
$|\psi_v\rangle=x|0\rangle+y|1\rangle$, one obtains
\begin{align}
	\ee(\rho_v;H_A)
	=
	\frac{2\kappa\epsilon_A|x|^2}{P_v}.
	\label{eq:appQubitEcomponent}
\end{align}

The two qubit Hamiltonian in Eq.~\eqref{eq:HAB} preserves parity and is therefore block diagonal in the even and odd parity sectors:
\begin{align}
	H_{AB}
	=
	\begin{aligned}[t]
		&\begin{pmatrix}
			S&0&0&J\\
			0&D&J&0\\
			0&J&-D&0\\
			J&0&0&-S
		\end{pmatrix}
		\simeq
		\begin{pmatrix}S&J\\J&-S\end{pmatrix}_{\rm even}
		\oplus
		\begin{pmatrix}D&J\\J&-D\end{pmatrix}_{\rm odd}.
	\end{aligned}
	\label{eq:appHmatrix}
\end{align}
Here, $\simeq$ denotes the basis reordering
$(|00\rangle,|11\rangle,|01\rangle,|10\rangle)$.
The even and odd parity blocks have eigenvalues
$\pm\Omega_+$ and $\pm\Omega_-$, respectively. Moreover,
\begin{align}
	\Omega_+^2-\Omega_-^2
	=
	S^2-D^2
	=
	4\epsilon_A\epsilon_B
	\geqslant0,
\end{align}
which gives $\Omega_+\geqslant\Omega_-$ and hence the energy ordering stated below Eq.~\eqref{eq:omegas}.

For a four-dimensional conditional state of the form in Eq.~\eqref{eq:appRankOne}, the isotropic part is unchanged by unitary rearrangements. The passive and active states therefore differ only in whether the excess eigenvalue weight $\Delta_v$ is assigned to the lowest or highest energy level. It follows that
\begin{align}
	\capc(\rho_v;H_{AB})
	&=
	2\Omega_+\Delta_v,
	\label{eq:appTwoQubitC}\\
	\ee(\rho_v;H_{AB})
	&=
	\Delta_v
	\left(
	\frac{\langle\psi_v|H_{AB}|\psi_v\rangle}{n_v}
	+\Omega_+
	\right).
	\label{eq:appTwoQubitE}
\end{align}
For $n_v=0$ and $P_v>0$, the conditional state is maximally mixed, and the ergotropy in Eq.~\eqref{eq:appTwoQubitE} is understood by continuous extension to be zero.
We now use Eqs.~\eqref{eq:appRankOneP}-\eqref{eq:appTwoQubitE} to derive the conditional quantities and constrained optima for the four resource states considered in the main text.

\subsection{Proof of the criterion for optimal product measurement outcomes}\label{app:criterion}

For $0<p<1$, we have
$G_p\geqslant \frac{p}{d_M}\1_M>0$.
With the change of variables $y=G_p^{1/2}z$, the conditional ergotropy quotient becomes
\begin{align}
	\frac{z^\dagger Qz}{z^\dagger G_pz}
	=
	\frac{y^\dagger G_p^{-1/2}QG_p^{-1/2}y}{y^\dagger y}
	\leqslant
	\lambda_p.
\end{align}
By the Rayleigh principle, equality holds if and only if $Qz=\lambda_pG_pz$, or equivalently $z\in\mathcal V_p$.

For any nonzero $z\in\mathcal V_p$, let $c=\overline z/\sqrt{z^\dagger z}$. The auxiliary measurement vector with coefficient column $c$ can be completed to an orthonormal basis and therefore realized as an outcome of a joint projective measurement. 
If $z$ is a product vector, so is $c$; completing each local factor of the corresponding measurement vector to a local orthonormal basis yields a product projective measurement containing that outcome. 
Conversely, if a product measurement outcome with coefficient column $c$ reaches the unrestricted joint optimum, then $\overline c\in\mathcal V_p$.

The generalized Rayleigh quotient is continuous on the compact set of normalized product vectors. 
Hence, if $\mathcal V_p$ contains no product vector, the maximum over product outcomes is attained strictly below $\lambda_p$. 
This proves the product attainability statement for conditional ergotropy in Prop.~\ref{prop:attainability}.

For $r\coloneqq z^\dagger Rz$ on the unit sphere, the success probability, conditional capacity, and probability-weighted capacity are
\begin{align}
	P_v
	&=
	\frac{p}{d_M}+(1-p)r,
	\label{eq:prob}\\[3pt]
	\capc(\rho_T^v;H_T)
	&=
	\frac{(1-p)\Delta E\,r}
	{p/d_M+(1-p)r},\\[4pt]
	Y_{\capc,v}
	&=
	(1-p)\Delta E\,r \label{eq:prob_capcity}.
\end{align}
Under the assumptions of the proposition, all three Eqs.~\eqref{eq:prob}-\eqref{eq:prob_capcity} are strictly increasing functions of $r$. 
Their maximizers are therefore precisely the unit vectors in the eigenspace of $R$ associated with its largest eigenvalue. 
Likewise, since
$Y_{\ee,v}=(1-p)z^\dagger Qz$,
its maximizers are the unit vectors in the maximal eigenspace of $Q$.

A single measurement outcome can maximize the success probability, conditional capacity, and conditional ergotropy simultaneously only if the maximal eigenspace of $R$ has a nonzero intersection with $\mathcal V_p$. 
This intersection is nontrivial for all four examples considered in the main text, but need not be so for a general resource state.
For two auxiliary qubits, write the coefficient vector $z$ as the $2\times2$ matrix
\begin{align}
	Z
	\coloneqq
	\begin{pmatrix}
		z_{00}&z_{01}\\
		z_{10}&z_{11}
	\end{pmatrix}.
\end{align}
A nonzero vector $z$ is a product vector if and only if $\det Z=0$. 
Now consider any complex subspace containing two linearly independent vectors $z_1$ and $z_2$, with corresponding coefficient matrices $Z_1$ and $Z_2$. 
The determinant $\det(sZ_1+tZ_2)$ is a homogeneous quadratic polynomial in the complex coefficients $s$ and $t$. 
It either vanishes identically or has a zero for some nonzero pair $(s,t)$.
Since $z_1$ and $z_2$ are linearly independent, the corresponding vector $sz_1+tz_2$ is nonzero and therefore provides a product vector in the subspace. 
Thus, every complex subspace of two qubits with dimension at least two contains a product vector, consistent with the general dimension bound of Ref.~\cite{Parthasarathy2004}.

It follows that the maximizing subspace $\mathcal V_p$ contains no product vector if and only if it is one-dimensional and spanned by an entangled vector, which gives the two qubit characterization stated after Prop.~\ref{prop:attainability}. 
The endpoints require separate treatment. 
At $p=0$, $G_p$ may be singular, so the argument based on positive definiteness no longer applies. 
Every conditional state with nonzero probability is then pure and has capacity $\Delta E$, while its ergotropy depends on the normalized projected state.
At $p=1$, $P_v=1/d_M$ and both energetic quantities vanish for every measurement outcome. 
In the derivations below, these endpoint cases are treated directly, without assigning a conditional state to any zero-probability outcome.

\subsection{Three qubit W state}

Projecting the pure W$_3$ component in Eq.~\eqref{eq:W3} onto the auxiliary outcome $|v\rangle$ defined in Eq.~\eqref{eq:v} gives
	\begin{align}
		(\1_A\otimes\langle v|)|\mathrm{W}_3\rangle
		&=
		\frac{1}{\sqrt3}
		\left[
		\overline{(c_{21}+c_{31})}|0\rangle_A
		+
		\overline{c_{11}}|1\rangle_A
		\right]
		=
		\frac{|w_v\rangle}{\sqrt3}.
	\end{align}
Here,
	$|w_v\rangle
	=
	\overline{(c_{21}+c_{31})}|0\rangle_A
	+
	\overline{c_{11}}|1\rangle_A$
	is the unnormalized pure component on battery $A$ introduced in Eq.~\eqref{eq:Wconditional}, with
	\begin{align}
		\langle w_v|w_v\rangle
		=
		|c_{21}+c_{31}|^2
		+
		|c_{11}|^2
		=
		N.
	\end{align}
The label $v$ refers to the selected auxiliary measurement outcome, and the overbar denotes complex conjugation of the corresponding measurement amplitudes. 
For $N>0$, the normalized projected pure component on the battery is $|w_v\rangle/\sqrt N$.

After projection onto the auxiliary outcome and tracing out the auxiliary subsystem, the maximally mixed component contributes $\frac p8\1_2$ to the battery branch. Equation~\eqref{eq:Wconditional} therefore takes the form of Eq.~\eqref{eq:appRankOne} with $|\psi_v\rangle=|w_v\rangle$ and
\begin{align}
	d=2,\quad
	\eta=\frac{p}{8},\quad
	\kappa=\frac{1-p}{3},\quad
	n_v=N.
\end{align}
Substituting these quantities into Eqs.~\eqref{eq:appRankOneP}, \eqref{eq:appQubitC}, and \eqref{eq:appQubitEcomponent} gives Eqs.~\eqref{eq:WP}-\eqref{eq:WE}.

Before any measurement on $BC$, the reduced battery state is
\begin{align}
	\rho_p^A
	=
	\frac{4-p}{6}|0\rangle\langle0|
	+
	\frac{2+p}{6}|1\rangle\langle1|.
\end{align}
Since the upper level $|0\rangle$ is more populated than the lower level $|1\rangle$ for $p<1$, the battery state is active. 
Its eigenvalue difference is $(1-p)/3$, and Eq.~\eqref{eq:W3before} follows immediately.

Let $s\coloneqq c_{21}+c_{31}$, $q\coloneqq|c_{11}|^2$, and $\zeta\coloneqq|s|^2$, so that $N=q+\zeta$. 
Using normalization together with the Cauchy-Schwarz inequality,
\begin{align}
	N
	&=
	|c_{11}|^2+|c_{21}+c_{31}|^2\\
	&\leqslant
	|c_{11}|^2
	+2\bigl(|c_{21}|^2+|c_{31}|^2\bigr)\\
	&=
	2-|c_{11}|^2-2|c_{41}|^2
	\leqslant2.
	\label{eq:appW3Nbound}
\end{align}
Saturation requires $c_{11}=c_{41}=0$ together with equality in the Cauchy-Schwarz bound.
Hence $c_{21}$ and $c_{31}$ must have the same phase and equal magnitude, with $|c_{21}|=|c_{31}|=1/\sqrt2$. 
Therefore, up to an overall phase, the unique measurement vector that maximizes $N$ is the Bell state in Eq.~\eqref{eq:W3optvector}.

Here and below, $\capc\coloneqq\capc(\rho_v;H)$ and $\ee\coloneqq\ee(\rho_v;H)$ denote the conditional capacity and ergotropy, with $(\rho_v,H)=(\rho_A^v,H_A)$ for a single-qubit battery and $(\rho_v,H)=(\rho_{AB}^v,H_{AB})$ for a two-qubit battery. The relevant derivatives are
\begin{align}
	\frac{\partial P_v}{\partial N}
	&=
	\frac{1-p}{3}
	\geqslant0,\\[3pt]
	\frac{\partial\capc}{\partial N}
	&=
	\frac{24p(1-p)}
	{\left[3p+4(1-p)N\right]^2}
	\epsilon_A
	\geqslant0,
	\label{eq:appW3PCderivatives}\\[4pt]
	\frac{\partial\ee}{\partial\zeta}
	&=
	\frac{8(1-p)\left[3p+4(1-p)q\right]}
	{\left[3p+4(1-p)(q+\zeta)\right]^2}
	\epsilon_A
	\geqslant0,\\[4pt]
	\frac{\partial\ee}{\partial q}
	&=
	-\frac{32(1-p)^2\zeta}
	{\left[3p+4(1-p)(q+\zeta)\right]^2}
	\epsilon_A
	\leqslant0.
	\label{eq:appW3Ederivatives}
\end{align}
For $0<p<1$, the conditional ergotropy is therefore maximized by maximizing $\zeta$ and minimizing $q$ over the allowed domain. 
This gives $\zeta=2$ and $q=0$, corresponding to the Bell outcome in Eq.~\eqref{eq:W3optvector}. 
The same outcome also maximizes $P_v$ and $\capc$, which proves Eq.~\eqref{eq:W3max}.

At $p=0$, every outcome with nonzero probability satisfies
$\ee=2\epsilon_A\zeta/(q+\zeta)$. 
The conditional ergotropy is therefore maximal for any measurement vector with $q=0$ and $\zeta>0$. 
Among these outcomes, however, only the Bell outcome, up to an overall phase, also maximizes the success probability and the corresponding probability weighted quantities. 
At the opposite endpoint, $p=1$, the postselected battery state is maximally mixed for every measurement outcome, $P_v=1/4$, and both the capacity and ergotropy vanish.

For a product measurement outcome, write
\begin{align}
	|v\rangle
	=
	(\alpha|0\rangle+\beta|1\rangle)
	\otimes
	(\gamma|0\rangle+\delta|1\rangle),
	\label{eq:appProductVector}
\end{align}
with both local factors normalized. 
The corresponding coefficients are
\begin{align}
	c_{11}=\alpha\gamma,\quad
	c_{21}=\alpha\delta,\quad
	c_{31}=\beta\gamma,\quad
	c_{41}=\beta\delta.
\end{align}

For fixed $x_\alpha\coloneqq|\alpha|^2$, the quantity $N$ can be written as the quadratic form
\begin{align}
	N
	=
	\begin{pmatrix}
		\overline\gamma&\overline\delta
	\end{pmatrix}
	\begin{pmatrix}
		1&\overline\beta\alpha\\
		\overline\alpha\beta&x_\alpha
	\end{pmatrix}
	\begin{pmatrix}
		\gamma\\
		\delta
	\end{pmatrix}.
\end{align}
Maximizing over the second local qubit therefore gives the largest eigenvalue of the $2\times2$ matrix,
\begin{align}
	f(x_\alpha)
	\coloneqq
	\frac{1+x_\alpha+\sqrt{1+2x_\alpha-3x_\alpha^2}}{2}.
\end{align}
The stationary point satisfies $f'(x_\alpha)=0$ at $x_\alpha=2/3$, where $f(2/3)=4/3$, while the endpoint values are smaller. 
Hence,
\begin{align}
	N_{\max}^{\rm prod}
	=
	\frac43.
	\label{eq:appW3ProductN}
\end{align}

One product outcome that attains this bound within Protocol I is
\begin{align}
	|v\rangle
	=
	\left(
	\sqrt{\frac23}|0\rangle
	+
	\frac{e^{i\chi}}{\sqrt3}|1\rangle
	\right)^{\otimes2},
\end{align}
where $\chi\in\mathbb R$ is arbitrary.
Substituting Eq.~\eqref{eq:appW3ProductN} into Eqs.~\eqref{eq:WP} and \eqref{eq:WC} gives the product optima for the success probability and conditional capacity in Eqs.~\eqref{eq:W3productP} and \eqref{eq:W3productC}.

To bound the conditional ergotropy over product measurement outcomes, first note that
\begin{align}
	|s|
	=
	|\alpha\delta+\beta\gamma|
	\leqslant
	\sqrt{|\alpha|^2+|\beta|^2}
	\sqrt{|\delta|^2+|\gamma|^2}
	=
	1.
	\label{eq:appW3ProductS}
\end{align}
Hence $\zeta=|s|^2\leqslant1$. 
For $0<p<1$, Eq.~\eqref{eq:appW3Ederivatives} further shows that saturation of the product ergotropy bound requires the smallest allowed value of $q=|\alpha\gamma|^2$, namely $q=0$.
The product outcomes $|01\rangle$ and $|10\rangle$ satisfy $\zeta=1$ and $q=0$, and therefore attain the bound. 
Substituting these values into Eq.~\eqref{eq:WE} gives Eq.~\eqref{eq:W3product}.
At $p=0$, the maximal conditional ergotropy is again $2\epsilon_A$, while at $p=1$ the ergotropy vanishes for every measurement outcome.

For the parametrization used in Fig.~\ref{fig:W3performance}, Eq.~\eqref{eq:Wx} gives $x=q+\zeta-1$. 
Since $\zeta\geqslant0$, we have $x\geqslant q-1$. 
On the other hand,
\begin{align}
	\zeta
	\leqslant
	2\bigl(|c_{21}|^2+|c_{31}|^2\bigr)
	\leqslant
	2(1-q),
\end{align}
which gives $x\leqslant1-q$. Both boundary values are attainable, and continuous variation of the measurement amplitudes fills the entire interval between them. 
For fixed $q$,
\begin{align}
	\frac{\partial\ee}{\partial x}
	=
	\frac{8(1-p)\left[3p+4(1-p)q\right]}
	{\left[3p+4(1-p)(1+x)\right]^2}
	\epsilon_A
	\geqslant0,
\end{align}
establishing the monotonic increase of the conditional ergotropy with $x$ used in Fig.~\ref{fig:W3performance}.

Here and below, we omit the outcome subscript $v$, writing $Y_{\capc}\coloneqq P_v\capc$ and $Y_{\ee}\coloneqq P_v\ee$. Finally, multiplying the conditional capacity and ergotropy by the success probability in Eq.~\eqref{eq:WP} removes the common normalization denominator and gives
\begin{align}
	Y_{\capc}
	&=
	\frac{2(1-p)N}{3}\epsilon_A,\\
	Y_{\ee}
	&=
	\frac{2(1-p)\zeta}{3}\epsilon_A.
\end{align}
The bounds in Eqs.~\eqref{eq:appW3Nbound}, \eqref{eq:appW3ProductN}, and \eqref{eq:appW3ProductS} then directly yield all probability weighted bounds stated below Eq.~\eqref{eq:W3yield}.

\subsection{Three qubit GHZ state}
	
Projecting the pure GHZ$_3$ component onto the selected auxiliary outcome $|v\rangle$ gives
	\begin{align}
		(\1_A\otimes\langle v|)|\mathrm{GHZ}_3\rangle
		=
		\frac{
			\overline{c_{11}}|0\rangle_A
			+
			\overline{c_{41}}|1\rangle_A
		}{\sqrt2}
		=
		\frac{|g_v\rangle}{\sqrt2}.
	\end{align}
Here,
	$|g_v\rangle
	=
	\overline{c_{11}}|0\rangle_A
	+
	\overline{c_{41}}|1\rangle_A$
is the unnormalized pure component on battery $A$ introduced in Eq.~\eqref{eq:GHZ3vector}, with
	\begin{align}
		\langle g_v|g_v\rangle
		=
		|c_{11}|^2+|c_{41}|^2
		=
		m.
	\end{align}
The label $v$ identifies the selected measurement outcome on $BC$; for $m>0$, the corresponding normalized projected pure component on the battery is $|g_v\rangle/\sqrt m$.

Equation~\eqref{eq:GHZ3state} therefore takes the form of Eq.~\eqref{eq:appRankOne} with $|\psi_v\rangle=|g_v\rangle$ and
\begin{align}
	d=2,\quad
	\eta=\frac{p}{8},\quad
	\kappa=\frac{1-p}{2},\quad
	n_v=m.
\end{align}
Substituting these quantities into Eqs.~\eqref{eq:appRankOneP}, \eqref{eq:appQubitC}, and \eqref{eq:appQubitEcomponent} gives Eqs.~\eqref{eq:GHZ3P}-\eqref{eq:GHZ3E}. Before any measurement on $BC$, tracing out the auxiliary pair gives
$\rho_p^A=\1_2/2$, so both the battery capacity and ergotropy vanish.

For the three qubit GHZ branch, set $q\coloneqq|c_{11}|^2$, so that $0\leqslant q\leqslant m\leqslant1$. 
Here, $q$ gives the weight of the upper battery level in the unnormalized pure component $|g_v\rangle$.
The success probability and conditional capacity are both nondecreasing functions of $m$, as seen from
\begin{align}
	\frac{\partial P_v}{\partial m}
	&=
	\frac{1-p}{2}
	\geqslant0,\\
	\frac{\partial\capc}{\partial m}
	&=
	\frac{4p(1-p)}{[p+2(1-p)m]^2}\epsilon_A
	\geqslant0.
	\label{eq:appGHZ3PCderivatives}
\end{align}
For fixed $m$, Eq.~\eqref{eq:GHZ3E} is linear and nondecreasing in $q$, and is therefore maximized at $q=m$. Along this boundary, the conditional ergotropy coincides with the capacity and is again nondecreasing in $m$. Hence, the common optimum occurs at $m=q=1$, corresponding to the product outcome $|v\rangle=|00\rangle$, which gives Eq.~\eqref{eq:GHZ3max}.

The probability-weighted capacity and ergotropy simplify to
\begin{align}
	Y_{\capc}
	&=
	(1-p)m\epsilon_A,\\
	Y_{\ee}
	&=
	(1-p)q\epsilon_A,
\end{align}
and are maximized by the same product outcome $|00\rangle$.

For the fixed $q$ curves in Fig.~\ref{fig:GHZ3performance}, we use $m=1-r$, with normalization imposing
$0\leqslant r\leqslant1-q$. Direct differentiation gives
\begin{align}
	\frac{\partial P_v}{\partial r}
	&=
	-\frac{1-p}{2}
	\leqslant0,\\
	\frac{\partial\capc}{\partial r}
	&=
	-\frac{4p(1-p)}
	{\left[p+2(1-p)(1-r)\right]^2}
	\epsilon_A
	\leqslant0,\\[4pt]
	\frac{\partial\ee}{\partial r}
	&=
	\frac{8(1-p)^2q}
	{\left[p+2(1-p)(1-r)\right]^2}
	\epsilon_A
	\geqslant0.
\end{align}
Thus, at fixed $q$, increasing the weight outside the even parity sector lowers both the success probability and the conditional capacity, while increasing the normalized conditional ergotropy. 
This increase comes entirely from the renormalization of the postselected branch, since the upper level weight $q$ is held fixed. 
The physical domain ends at $r=1-q$; values beyond this boundary are incompatible with a normalized measurement vector and therefore play no role in the global optimization.

At $p=0$, every nonzero probability branch has conditional capacity $2\epsilon_A$, while the ergotropy is $2q\epsilon_A/m$. The outcome $|00\rangle$ still maximizes the success probability and both probability-weighted quantities.
At $p=1$, every postselected battery state is maximally mixed, and both the capacity and ergotropy vanish.


\subsection{Four qubit GHZ state}
\label{app:GHZ4product}

Before any measurement on the auxiliary pair $CD$, tracing out $CD$ gives the reduced battery state
\begin{align}
	\rho_p^{AB}
	=
	\frac{1-p}{2}
	\bigl(
	|00\rangle\langle00|
	+
	|11\rangle\langle11|
	\bigr)
	+
	\frac{p}{4}\1_4.
	\label{eq:appGHZ4Reduced}
\end{align}
Its spectrum consists of two larger eigenvalues $(2-p)/4$ and two smaller eigenvalues $p/4$. 
The mean energy is zero because the contributions of $|00\rangle$ and $|11\rangle$ cancel. 
The corresponding passive energy is
$-(1-p)(\Omega_++\Omega_-)/2$, while the active passive energy difference is
$(1-p)(\Omega_++\Omega_-)$. 
These relations give the premeasurement capacity and ergotropy in Eq.~\eqref{eq:GHZ4before}.

Projecting the pure GHZ$_4$ component onto the selected auxiliary outcome $|v\rangle$ gives
\begin{align}
	(\1_{AB}\otimes\langle v|)|\mathrm{GHZ}_4\rangle
	&=
	\frac{
		\overline{c_{11}}|00\rangle_{AB}
		+
		\overline{c_{41}}|11\rangle_{AB}
	}{\sqrt2}=
	\frac{|G_v\rangle}{\sqrt2}.
\end{align}
Here,
$|G_v\rangle
=
\overline{c_{11}}|00\rangle_{AB}
+
\overline{c_{41}}|11\rangle_{AB}$
is the unnormalized pure component on battery $AB$ introduced in Eq.~\eqref{eq:GHZ4state}, with
\begin{align}
	\langle G_v|G_v\rangle
	=
	|c_{11}|^2+|c_{41}|^2
	=
	m.
\end{align}
The label $v$ identifies the selected measurement outcome on $CD$. 
We use $|G_v\rangle$ here to distinguish the two qubit battery component from $|g_v\rangle$ in the three qubit GHZ case, which belongs to the single qubit battery $A$.

The conditional state in Eq.~\eqref{eq:GHZ4state} therefore takes the form of Eq.~\eqref{eq:appRankOne} with $|\psi_v\rangle=|G_v\rangle$ and
\begin{align}
	d=4,\quad
	\eta=\frac{p}{16},\quad
	\kappa=\frac{1-p}{2},\quad
	n_v=m.
\end{align}
For the four qubit GHZ branch, define the common normalization denominator
$d_4\coloneqq4P_v=p+2(1-p)m$, where the subscript $4$ labels the four-qubit resource. Eqs.~\eqref{eq:appRankOneP}, \eqref{eq:appDelta}, and \eqref{eq:appTwoQubitC} then give
\begin{align}
	P_v
	&=
	\frac{d_4}{4},\\
	\Delta_v
	&=
	\frac{2(1-p)m}{d_4},\\
	\capc
	&=
	\frac{4(1-p)m\Omega_+}{d_4}.
\end{align}

For $m>0$, the normalized pure component of the postselected battery branch is
\begin{align}
	|\widehat G_v\rangle
	\coloneqq
	\frac{|G_v\rangle}{\sqrt m}
	=
	\sqrt t\,e^{-i\arg c_{11}}|00\rangle_{AB}
	+
	\sqrt{1-t}\,e^{-i\arg c_{41}}|11\rangle_{AB},
\end{align}
where $t=|c_{11}|^2/m$ is the population of $|00\rangle_{AB}$ and
$\phi=\arg(c_{11})-\arg(c_{41})$ is the relative phase between the two even-parity components.
At $t=0$ or $1$, the phase of the vanishing component is irrelevant.  
Using the even parity block of Eq.~\eqref{eq:appHmatrix}, its energy expectation is
\begin{align}
	\langle\widehat G_v|H_{AB}|\widehat G_v\rangle
	=
	(2t-1)S
	+
	2\sqrt{t(1-t)}J\cos\phi.
	\label{eq:appGHZ4Energy}
\end{align}
Substituting this expression into Eq.~\eqref{eq:appTwoQubitE} gives the conditional ergotropy in Eq.~\eqref{eq:GHZ4E}.

Set $\xi\coloneqq2t-1$, so that
$2\sqrt{t(1-t)}=\sqrt{1-\xi^2}$. The factor controlling the ergotropy in Eq.~\eqref{eq:GHZ4E} can then be written as
\begin{align}
	\mathcal B(\xi,\phi)
	\coloneqq
	S\xi
	+
	J\sqrt{1-\xi^2}\cos\phi
	+
	\Omega_+.
\end{align}
By the Cauchy-Schwarz inequality,
\begin{align}
	\left|S\xi
	+
	J\sqrt{1-\xi^2}\cos\phi
	\right|
	\leqslant
	\sqrt{S^2+J^2}
	\sqrt{
		\xi^2+(1-\xi^2)\cos^2\phi}\leqslant	\Omega_+.
\end{align}
Hence,
$0\leqslant\mathcal B\leqslant2\Omega_+$.
For $\Omega_+>0$, the upper bound is attained for
\begin{align}
	\cos\phi
	&=
	1,\quad
	\xi
	=
	\frac{S}{\Omega_+},\quad
	t
	=
	t_\star
	=
	\frac12
	\left(
	1+\frac{S}{\Omega_+}
	\right),
\end{align}
which gives Eq.~\eqref{eq:GHZ4t}. For $J>0$, these relative-amplitude and phase conditions are required for saturation, up to an overall phase. When $J=0$, one has $t_\star=1$, and the relative phase becomes irrelevant.

The dependence on the branch weight $m$ is determined by
\begin{align}
	\frac{\partial P_v}{\partial m}
	&=
	\frac{1-p}{2}
	\geqslant0,\\
	\frac{\partial\capc}{\partial m}
	&=
	\frac{4p(1-p)\Omega_+}{d_4^2}
	\geqslant0,\\
	\frac{\partial\ee}{\partial m}
	&=
	\frac{2p(1-p)\mathcal B}{d_4^2}
	\geqslant0.
	\label{eq:GHZ4dm}
\end{align}
Thus, the common optimum is obtained at $m=1$ together with
$\mathcal B=2\Omega_+$, yielding Eq.~\eqref{eq:GHZ4max}.

The enhancement relative to the premeasurement battery state follows from
\begin{align}
	\frac{4\Omega_+}{2-p}
	\geqslant
	2\Omega_+
	\geqslant
	\Omega_++\Omega_-.
	\label{eq:appGHZ4CapacityComparison}
\end{align}
For $\Omega_+>0$, equality in the first inequality occurs only at $p=0$, while equality in the second occurs only when $\Omega_+=\Omega_-$. 
The optimal conditional ergotropy is strictly larger than its premeasurement value for $p<1$ and $\Omega_+>0$, since
\begin{align}
	\frac{4\Omega_+}{2-p}
	\geqslant
	2\Omega_+
	>
	\frac{\Omega_++\Omega_-}{2}.
\end{align}

We next ask whether the joint ergotropy optimum can be attained by a product measurement outcome. 
For $0<p<1$, saturation at $\mathcal B=2\Omega_+>0$ makes the last derivative in Eq.~\eqref{eq:GHZ4dm} strictly positive, so the optimum requires $m=1$. 
Normalization then forces $c_{21}=c_{31}=0$. A two qubit measurement vector is product if and only if
$c_{11}c_{41}=c_{21}c_{31}$; hence, for $m=1$, product structure requires $c_{11}c_{41}=0$, or equivalently $t\in\{0,1\}$. For $J>0$, however, $\Omega_+>S$ and therefore $0<t_\star<1$. 
The maximizing outcome in Eq.~\eqref{eq:GHZ4vector} is thus entangled. 
Since the set of normalized product vectors is compact and the conditional ergotropy is continuous on it, the product optimum is strictly below the unrestricted joint optimum for $J>0$ and $0<p<1$.


Let $q\coloneqq|c_{11}|^2$, $r\coloneqq|c_{41}|^2$, and $m=q+r>0$. For a product measurement vector, the determinant condition together with normalization gives
\begin{align}
	1-m
	&=
	|c_{21}|^2+|c_{31}|^2\geqslant
	2|c_{21}c_{31}|
	=
	2\sqrt{qr}.
\end{align}
Writing $t\coloneqq q/m$ and
$h\coloneqq2\sqrt{t(1-t)}$, this constraint becomes
$m\leqslant(1+h)^{-1}$. The bound is attainable by a product vector with identical real local factors, while the relative phase can be chosen to maximize the interaction contribution.

Since $S\geqslant0$, exchanging $q$ and $r$ if necessary shows that an optimal product outcome may be chosen with $t\geqslant1/2$. We therefore parametrize
\begin{align}
	2t-1
	&=
	\cos\theta,\quad
	h
	=
	\sin\theta,
	\quad
	0\leqslant\theta\leqslant\frac{\pi}{2}.
\end{align}
For fixed $t$, the monotonicity in Eq.~\eqref{eq:GHZ4dm} allows us to take the largest admissible value of $m$. 
The same choice is valid at $p=0$, where the conditional ergotropy is independent of $m$. 
The product optimization therefore reduces to the single variable problem
\begin{align}
	\ee_{\max}^{\mathrm{prod}}
	&=
	2(1-p)
	\max_{0\leqslant\theta\leqslant\pi/2}
	\frac{\Omega_+
		+
		S\cos\theta
		+
		J\sin\theta}{2-p+p\sin\theta}.
	\label{eq:GHZ4productQuotient}
\end{align}
For any maximizing value of $\theta$, an optimal product measurement outcome can be chosen as $|\ell_t\rangle^{\otimes2}$, where
\begin{align}
	|\ell_t\rangle
	\coloneqq
	\frac{
		t^{1/4}|0\rangle
		+
		(1-t)^{1/4}|1\rangle
	}{
		\sqrt{\sqrt t+\sqrt{1-t}}
	},
	\quad
	t
	=
	\frac{1+\cos\theta}{2}.
\end{align}

For $p<1$, let $k\coloneqq2-p$. The sign of the derivative of the quotient in Eq.~\eqref{eq:GHZ4productQuotient} is determined by
\begin{align}
	d(\theta)
	\coloneqq
	(kJ-p\Omega_+)\cos\theta
	-
	kS\sin\theta
	-
	pS.
\end{align}
If
$kJ\leqslant p(\Omega_++S)$,
then $d(\theta)\leqslant0$ throughout
$0\leqslant\theta\leqslant\pi/2$, and the maximum occurs at the boundary $\theta=0$. 
Otherwise, $d(0)>0$. For $S>0$, $d(\theta)$ then decreases across the interval and has a unique zero, which determines the interior optimum.

Let $f$ denote the value of the quotient at this interior stationary point. 
The stationarity condition gives
\begin{align}
	kf
	=
	\Omega_+
	+
	\sqrt{S^2+(J-pf)^2}.
\end{align}
Using $\Omega_+^2=S^2+J^2$ then yields
\begin{align}
	f=\frac{k\Omega_+-pJ}{2(1-p)}.
\end{align}
The boundary condition
$kJ\leqslant p(\Omega_++S)$
is equivalent to $p\geqslant p_c$. 
Combining the boundary and interior branches therefore gives Eq.~\eqref{eq:GHZ4productE}.
At $p=1$, all energetic quantities vanish. When $J=0$, one has $p_c=0$, and the product outcome $|00\rangle$ is optimal for every noise fraction. 
When $S=0$ and $J>0$, one has $p_c=1$; the optimum occurs at $\theta=\pi/2$, giving
\begin{align}
	\ee_{\max}^{\mathrm{prod}}
	=
	2(1-p)J,
\end{align}
attained by $|+\rangle^{\otimes2}$, where
$|+\rangle\coloneqq(|0\rangle+|1\rangle)/\sqrt2$.

For the success probability weighted ergotropy, the bound
\begin{align}
	(\Omega_++S)(1+\sin\theta)
	-
	\bigl(\Omega_++S\cos\theta+J\sin\theta\bigr)
	=
	S(1-\cos\theta)
	+
	(\Omega_++S-J)\sin\theta\geqslant0
\end{align}
together with
$m\leqslant(1+\sin\theta)^{-1}$
gives Eq.~\eqref{eq:GHZ4productYield}. 
The bound is saturated at $\theta=0$, corresponding to the product outcome $|00\rangle$. 
Hence, even within the product class, the outcome that maximizes the conditional ergotropy need not maximize its success probability weighted contribution.


\subsection{Four qubit W state}

Tracing out the auxiliary pair $CD$ from the pure W$_4$ resource gives
\begin{align}
	\Tr_{CD}|\mathrm{W}_4\rangle\langle \mathrm{W}_4|
	=
	\frac12|00\rangle\langle00|
	+
	\frac12|\psi_+\rangle\langle\psi_+|,
	\quad
	|\psi_+\rangle
	=
	\frac{|01\rangle+|10\rangle}{\sqrt2}.
	\label{eq:appW4ReducedPure}
\end{align}
After admixing global white noise, the reduced battery state has two larger eigenvalues $(2-p)/4$ and two smaller eigenvalues $p/4$. 
The two pure components have energy expectations
\begin{align}
	\langle00|H_{AB}|00\rangle
	&=
	S,\\[4pt]
	\langle\psi_+|H_{AB}|\psi_+\rangle
	&=
	J.
\end{align}
The mean battery energy is therefore $(1-p)(S+J)/2$, while the passive energy is
$-(1-p)(\Omega_++\Omega_-)/2$. 
These relations give the premeasurement capacity and ergotropy in Eqs.~\eqref{eq:W4beforeC} and \eqref{eq:W4beforeE}.

Projecting the pure W$_4$ component onto the selected auxiliary outcome $|v\rangle$ gives
\begin{align}
	(\1_{AB}\otimes\langle v|)|\mathrm{W}_4\rangle
	&=
	\frac12
	\left[
	\overline{(c_{21}+c_{31})}|00\rangle_{AB}
	+
	\overline{c_{11}}
	\bigl(|01\rangle_{AB}+|10\rangle_{AB}\bigr)
	\right]=
	\frac{|W_v\rangle}{2}.
\end{align}
Here, $|W_v\rangle$ is the unnormalized pure component on the two qubit battery $AB$ introduced in Eq.~\eqref{eq:W4vector},
\begin{align}
	&|W_v\rangle
	=
	\overline{(c_{21}+c_{31})}|00\rangle_{AB}
	+
	\overline{c_{11}}
	\bigl(|01\rangle_{AB}+|10\rangle_{AB}\bigr),\\[4pt]
	&\langle W_v|W_v\rangle
	=
	|c_{21}+c_{31}|^2
	+
	2|c_{11}|^2
	=
	m.
\end{align}
The label $v$ identifies the selected measurement outcome on $CD$; for $m>0$, the corresponding normalized projected pure component on the battery is $|W_v\rangle/\sqrt m$. 
We use $|W_v\rangle$ here to distinguish this two-qubit battery component from the single qubit vector $|w_v\rangle$ introduced for the W$_3$ resource.

The conditional state in Eq.~\eqref{eq:W4state} therefore takes the form of Eq.~\eqref{eq:appRankOne} with $|\psi_v\rangle=|W_v\rangle$ and
\begin{align}
	d=4,\quad
	\eta=\frac{p}{16},\quad
	\kappa=\frac{1-p}{4},\quad
	n_v=m.
\end{align}
For the four qubit W branch, define the common normalization denominator
$d_{\mathrm{W}}\coloneqq4P_v=p+(1-p)m$. 
The upright subscript $\mathrm{W}$ distinguishes this scalar quantity from the battery vector $|W_v\rangle$ defined above. 
The corresponding excess eigenvalue weight is
\begin{align}
	\Delta_v
	=
	\frac{(1-p)m}{d_{\mathrm{W}}}.
\end{align}
Substituting this into Eq.~\eqref{eq:appTwoQubitC} gives the conditional capacity in Eq.~\eqref{eq:W4C}.

For the unnormalized vector $|W_v\rangle$, direct evaluation gives
\begin{align}
	\langle W_v|H_{AB}|W_v\rangle
	=
	S|c_{21}+c_{31}|^2
	+
	2J|c_{11}|^2
	=
	mS+u(J-S).
	\label{eq:appW4PureEnergy}
\end{align}
Combining Eq.~\eqref{eq:appW4PureEnergy} with Eq.~\eqref{eq:appTwoQubitE} then gives the conditional ergotropy in Eq.~\eqref{eq:W4E}.
We next determine the exact domain accessible to normalized auxiliary measurement vectors. 
By definition, $u\geqslant0$ and
$m-u=|c_{21}+c_{31}|^2\geqslant0$.
Moreover,
\begin{align}
	m
	=
	2|c_{11}|^2
	+
	|c_{21}+c_{31}|^2
	\leqslant
	2|c_{11}|^2
	+
	2\bigl(|c_{21}|^2+|c_{31}|^2\bigr)
	=
	2\bigl(1-|c_{41}|^2\bigr)
	\leqslant
	2.
	\label{eq:appW4DomainBound}
\end{align}
Hence, every normalized measurement vector gives a parameter pair satisfying
$0\leqslant u\leqslant m\leqslant2$.

Conversely, every point in this triangular region is physically attainable. 
One explicit realization is provided by the normalized real coefficients
\begin{align}
	c_{11}
	&=
	\sqrt{\frac{u}{2}},\qquad
	c_{21}
	=
	c_{31}
	=
	\frac{\sqrt{m-u}}{2},\qquad
	c_{41}
	=
	\sqrt{1-\frac{m}{2}}.
	\label{eq:appW4DomainConstruction}
\end{align}
Hence, the triangular region $0\leqslant u\leqslant m\leqslant2$ coincides exactly with the parameter domain generated by normalized auxiliary measurement vectors.

The success probability and conditional capacity are nondecreasing functions of $m$:
\begin{align}
	\frac{\partial P_v}{\partial m}
	&=
	\frac{1-p}{4}
	\geqslant0,\\
	\frac{\partial\capc}{\partial m}
	&=
	\frac{2p(1-p)\Omega_+}{d_{\mathrm{W}}^2}
	\geqslant0.
	\label{eq:appW4PCderivatives}
\end{align}
At fixed $m$, the conditional ergotropy is affine in $u$, with
\begin{align}
	\frac{\partial\ee}{\partial u}
	=
	\frac{(1-p)(J-S)}{d_{\mathrm{W}}}.
	\label{eq:appW4Uderivative}
\end{align}
The optimal parity component is therefore selected by the sign of $J-S$: the ergotropy is maximized at $u=0$ for $J<S$, at $u=m$ for $J>S$, and is independent of $u$ when $J=S$.

Along the boundary $u=0$,
\begin{align}
	\ee_{u=0}(m)
	&=
	\frac{(1-p)m(S+\Omega_+)}{d_{\mathrm{W}}},\\
	\frac{d\ee_{u=0}}{dm}
	&=
	\frac{p(1-p)(S+\Omega_+)}
	{d_{\mathrm{W}}^2}
	\geqslant0,
\end{align}
whereas along $u=m$,
\begin{align}
	\ee_{u=m}(m)
	&=
	\frac{(1-p)m(J+\Omega_+)}{d_{\mathrm{W}}},\\
	\frac{d\ee_{u=m}}{dm}
	&=
	\frac{p(1-p)(J+\Omega_+)}
	{d_{\mathrm{W}}^2}
	\geqslant0.
\end{align}
For $0<p<1$, these derivatives are strictly positive in every nontrivial case, so the optimum lies at the maximal allowed value $m=2$. 
The construction in Eq.~\eqref{eq:appW4DomainConstruction} then identifies $(u,m)=(0,2)$ with the Bell outcome and $(u,m)=(2,2)$ with the product outcome $|00\rangle$. 
This gives the joint optima in Eqs.~\eqref{eq:W4PCmax} and \eqref{eq:W4Emax}.

For $J<S$ and $0<p<1$, attaining the maximal ergotropy requires $u=0$ and $m=2$.
Equality in Eq.~\eqref{eq:appW4DomainBound} then requires $c_{11}=c_{41}=0$ and $c_{21}=c_{31}$, with an arbitrary common phase.
The resulting Bell vector violates the product condition $c_{11}c_{41}=c_{21}c_{31}$, establishing a strict gap between the product and joint optima.
At $p=0$, the conditional ergotropy on either optimizing boundary is independent of $m$ for $m>0$.
In this pure state limit, for $J<S$, the product vector $|01\rangle$ has $(u,m)=(0,1)$ and attains the same conditional capacity and ergotropy as the Bell vector, but its success probability is $1/4$, compared with $1/2$ for the Bell vector.
At $p=1$, all energetic quantities vanish.
For $J\geqslant S$, $|00\rangle$ is an optimal product vector; at $J=S$, every vector with $m=2$ attains the maximal ergotropy.

The success-probability-weighted ergotropy is
\begin{align}
	Y_{\ee}
	=
	\frac{1-p}{4}
	\left[
	m(S+\Omega_+)
	+
	u(J-S)
	\right].
\end{align}
Evaluating this expression at the two optimizing vertices gives the weighted optima stated below Eq.~\eqref{eq:W4Emax}.

We finally compare the optimized conditional quantities with their premeasurement values. 
For the capacity, the result follows directly from the inequality chain in Eq.~\eqref{eq:appGHZ4CapacityComparison}. 
For the ergotropy, when $J\leqslant S$,
\begin{align}
	S+\Omega_+-(J+\Omega_-)
	&=
	(S-J)+(\Omega_+-\Omega_-)
	\geqslant0,
\end{align}
whereas for $J\geqslant S$,
\begin{align}
	J+\Omega_+-(S+\Omega_-)
	&=
	(J-S)+(\Omega_+-\Omega_-)
	\geqslant0.
\end{align}
Together with $2/(2-p)\geqslant1$, these inequalities show that the maximal conditional ergotropy is no smaller than the corresponding premeasurement value, as stated in the main text.

\twocolumngrid
\bibliography{references}
\end{document}